\documentclass[pre, twocolumn, preprintnumbers, superscriptaddress]{revtex4-1}
\usepackage[]{graphicx}
\usepackage{epstopdf}
\usepackage{amssymb}
\usepackage{bm}
\usepackage{framed}
\usepackage{hyperref}

\begin{document}
\title{Finite-dimensional signature of spinodal instability in an athermal hysteretic transition}
\author{Anurag Banerjee}
\affiliation{Department of Physics, Ben-Gurion University of the Negev, Beer-Sheva 84105, Israel}
\author{Tapas Bar}\email{tapas.bar@icn2.cat}
\affiliation{ICN2-Institut Catal\`{a} de Nanoci\`{e}ncia i Nanotecnologia (CERCA-BIST-CSIC), Campus Universitat Aut\`{o}noma de Barcelona, 08193 Bellaterra, Barcelona, Spain}
\date{\today}
\begin{abstract}
We study the off-equilibrium critical phenomena across a hysteretic first-order transition in disordered athermal systems. The study focuses on the zero temperature random field Ising model (ZTRFIM) above the critical disorder for spatial dimensions $d=2,3,$ and $4$. We use Monte Carlo simulations to show that disorder suppresses critical slowing down in phase ordering time for finite-dimensional systems. The dynamic hysteresis scaling, the measure of explicit finite-time scaling, is used to subsequently quantify the critical slowing down. The scaling exponents in all dimensions increase with disorder strength and finally reach a stable value where the transformation is no longer critical. The associated critical behavior in the mean-field limit is very different, where the exponent values for various disorders in all dimensions are similar. The non-mean-field exponents asymptotically approach the mean-field value ($\Upsilon \approx 2/3$) with increase in dimensions. The results suggest that the critical features in the hysteretic metastable phase are controlled by inherent mean-field spinodal instability that gets blurred by disorder in low-dimension athermal systems.
\end{abstract}
\preprint{Physical Review B {\bf 107}, 024103 (2023)}
\maketitle

\section{Introduction}
\begin{figure}[!t]
\center
\includegraphics[scale=0.3]{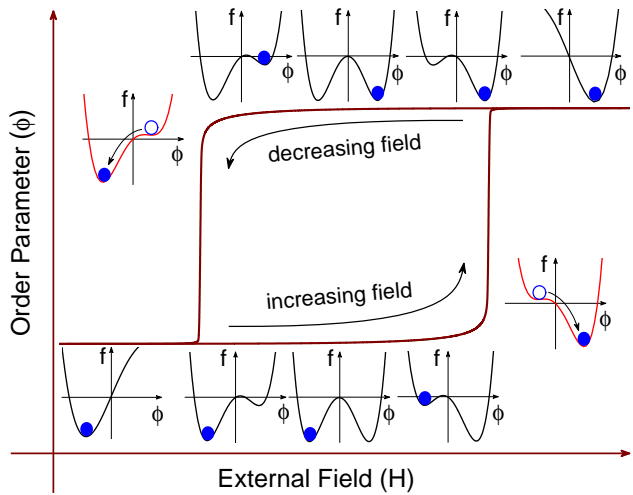}
\caption{A schematic diagram for the spinodal transitions. The order parameter $\phi$ and corresponding free-energy diagrams ($f-\phi$ curve) are exhibited as a function of increasing and decreasing fields. The $f-\phi$ curves in the middle represent the binodal points where the two minima are equal, and red $f-\phi$ diagrams are the two spinodals points (limit of metastability) where the double-well free energy switches to a single well, which is a conventional manifestation of continuous transitions. The system exhibits spinodal transition when the thermal fluctuations are insignificant to cross the free energy activation barrier between binodal and spinodal points.}
\label{Free_eng}
\center
\end{figure}

The critical-like features in abrupt hysteretic transition have recently been observed in various materials including transition metal oxide \cite{Basov_Nat18, Basov_Nat17, Bar_prl18, Kundu_prl20}, metal alloys \cite{Chandni_prl09, Bar_prb21}, martensitic transformation \cite{Keim_rmp19, Toth_prb14, Gallardo_prb10}, functional materials \cite{Book_Disorder}, amorphous solids \cite{Parisi_PNAS17, Tarjus_PNAS18, Nandi_prl16}, microbiology, and social, economic, climate, and other complex systems \cite{Scheffer_nat09, Scheffer_sci12}. Such ``surprising'' \cite{Keim_rmp19} behavior is not normal in terms of typical first-order phase transition formalism. Some of such transitions have been explained in terms of classical spinodal instability, a limiting point of metastability (Fig. \ref{Free_eng}), where the system behaves like a mean-field \cite{Binder_pra84, Bar_prl18, Kundu_prl20, Zapperi_prl97}. The stability of the metastable phase depends on the competition of disorder, thermal fluctuation, and activation barriers separating the two phases \cite{Book_Disorder}. Any fluctuations, linked with disorder or thermal, in the abrupt transition initiate nucleations before the extreme limit of metastability \cite{PTBook_Metastable}. In the long-range interacting system, thermal fluctuations are suppressed \cite{Binder_pra84, Planes_prl01}, and the metastable phase of the system approaches the spinodal point after multiple cycling of the materials (training) across the transition \cite{Francisco_prb04, Bar_prl18}. The divergence of correlation length and relaxation time scale (spinodal slowing down) signals the instability in experiments \cite{Basov_Nat17, Kundu_prl20, Bar_prb21}. The mean-field spinodal universality in disorder material might be explained in terms of training-induced self-organized criticality \cite{Francisco_prl07, Francisco_prb16}. However, the critical exponents often vary widely from mean-field predictions \cite{Liu_prl16} [see Table \ref{table_exp}] and therefore remains unexplained. In general, the training cannot tune the quenched disorders such as domain walls, friction, defects due to an underlying heterogeneous substrate, pinning defects, and kinetically arrested heterogeneity. Therefore, the correlation length of the system would be bounded by the local disorder points, and heterogeneous nucleation sites start to emerge before approaching the spinodal \cite{Cao_prb90, Imry_prb79, Fan_prb11, Scheifele_pre13, Wang_pre07}. As a result, a suppressed spinodal slowing down associated with a mild finite-size effect is expected to be observed \cite{Bar_prb21, Nandi_prl16, Bhowmik_pre19} that may explain such non-mean-field critical exponents. In this article, we investigate spinodal instability using a random field Ising model (RFIM) in the presence of quench disorder and under athermal conditions. The athermal (zero temperature) model mimics the fluctuationless kinetics associated with long-ranged potential, whereas the short-ranged Ising model only deals with the interplay of disorder and metastable barrier.

In RFIM, the critical signature in hysteretic transition has generally been observed in two distinct aspects: steady-state (slow-driven or quasistatic) and off-equilibrium (highly-driven). The steady-state studies are limited to the avalanche distribution and can explain the disorder-induced critical transition near the critical disorder \cite{Sethna_prl93, Sethna_prl95}. Away from the critical point, the power-law behavior of avalanche distribution is not adequately understood \cite{Sethna_prl95}. One study attempts to explain such phenomena at a low disorder regime in the context of spinodal instability \cite{Nandi_prl16}. However, most of the hysteretic transitions in materials are not single-step processes; instead they show a broad transition accompanied by return point memory indicating the disorder in the system is greater than the critical disorder \cite{Sethna_prl93, Keim_rmp19, Pierce_prb07}. Therefore, further investigations are required above the critical point.  On the other hand, the off-equilibrium aspect of critical phenomena such as dynamic hysteresis scaling and phase ordering dynamics are comparatively easy to measure in experiments. Not surprisingly, numerous assessments have been reported for different materials \cite{Lee_pre16, Yildiz_pre04, Wang_JCP11, He_prl93, Jiang_prb95, Pan_apl03, Liu_jap99, Kim_prb97, Jung_prl90, Bar_prb21, Bar_prl18, Kuang_ssc00, Wongdamnern_jap09, Wongdamnern_mcp10, wongdamnern_kem10, Yimnirun_apl06, Yimnirun_apl07, Zhang_ssc96, Pan_msc03, Prajapati_NJP22}. In theory, several attempts have also been made in diversified models, but the results are often inconsistent with one another (except in the mean-field limit). Such studies are designed to describe specific experimental result \cite{Rao_prb90, Rao_prb91, RaoComment_prl92, Zhong_FrontPhys17_Temp, Zhong_FrontPhys17_Field, Zhong_prl95, Zhong_prl05, Shukla_pre18}. Therefore, the origin of this general phenomenon is not properly explored. In this work, we systematically study the off-equilibrium critical phenomena from a general perspective that describes a large class of the experimentally reported dynamical critical exponents in various systems.

\section{The Model and Simulation}
We consider a $d$-dimensional ($d= 2, 3, 4$ ) random field Ising model in which every spin interacts with its nearest neighbors. A random field added to an external field acts as a disorder of the system. The Hamiltonian of the model read as
\begin{equation}\label{eqn:Hamiltonian}
\mathcal{H} = -J\sum\limits_{\langle i,j\rangle} s_is_j - \sum\limits_{i} [H(t)+h_i]s_i \ ,
\end{equation}
where $J$ is the nearest-neighbor coupling strength of Ising spins $s_i$,  $s_i = \pm 1$, placed on the $d$-dimensional hypercubic lattice of system of linear size $L$. The spin interacts ferromagnetically with strength $J = 1$ under the periodic boundary condition. A time-dependent spatially uniform external field, $H(t)$, and a time-independent but site-dependent random field, $h_i$ is applied. The random field $h_i$ is taken from a Gaussian distribution, $V(h)$,
\begin{equation}
V(h) = \frac{1}{\sqrt{2 \pi \sigma^2}} e^{-h^2/(2\sigma^2)},
\label{eqn:Disorder}
\end{equation}
where the width of the distribution represents the disorder strength of a single realization. We present all the physical quantities after averaging over a sufficient number of independent disorder realizations ($\sim 20-500$). Since we are interested in the athermal system, the thermal fluctuation in the model can be neglected by performing zero-temperature simulations. Therefore the spin-flip is completely determined by the sign change of the local field at each site \cite{Sethna_prl93}, 
\begin{equation}
E_i=J\sum_j s_j+ h_i+H.
\label{eqn:flip}
\end{equation}

The zero temperature random field Ising model (ZTRFIM) shows an external field-dependent hysteretic magnetic transition (or switching) for a large range of disorder values $\sigma$. The transition could be a single or multiple-step (avalanche) process depending upon the strength of the disorder. There is a critical disorder, $\sigma = \sigma_c$, above which single-step transition never happens. At $\sigma = \sigma_c$, the avalanche of all sizes exists that follows a long (several decades) power law size distribution connected to a disorder-induced continuous transition  (we will say this is classical-critical point to avoid ambiguity) \cite{Sethna_prl93, Sethna_prl95}. Here, we focus on critical-like field-induced transitions for $\sigma \geq \sigma_c$.

\subsection{Phase ordering dynamics}
\label{Phase-ordering}
We perform the phase ordering dynamics of the ZTRFIM on a $d$-dimensional lattice. We start with a system of fully polarized spins and suddenly tune the magnetic field close to the coercive field, the field at which the magnetization reverses. We study the time required to reach the steady state after the quench. During this interval, the system goes through successive set of spin-flips and finally arrives at a steady state. Such phase ordering (or continuous ordering) is generally measured through quench-and-hold experiments \cite{Bar_prl18, Bar_prb21, Bray_AP02}. We extract the relaxation time constant from the temporal evolution of the net magnetization. The details algorithm is presented below.

\begin{enumerate}
\item The spin at every site is either up or down ($s_i= 1$ or $s_i = -1$) depending upon the sign of the initial field $H_0$.
\item We quench the external magnetic field to $H = H_f$ at time $t = 0$ and check if the local field, defined in Eq.~\ref{eqn:flip}, changes sign on any site.
\item If there is a sign change of the local field for at least one site, we flip the spins on those sites in the next time step $t = t+1$.
\item We check all the sites and repeat step 3 until no site changes the sign of the local field, which indicates the system has reached a steady state.
\item The time required to reach this is considered as the phase ordering times (or relaxation time) $\tau$ for that particular quenched field $H_f$.
\item We continue this process (steps 1-5) for different quench fields $H = H_f$  to get phase ordering times throughout the transition regions both for increasing and decreasing fields.
\end{enumerate}

\begin{figure}[!t]
\center
\includegraphics[scale=0.31]{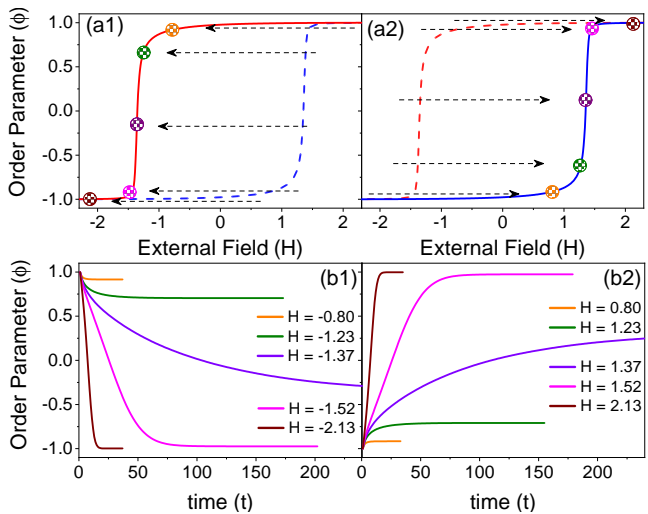}
\caption{The phase ordering simulations demonstrate for specified quenched external fields ($\otimes$) after decreasing (a1) and increasing (a2) field-quenched from the complete spin-polarized states. (b1), (b2) the corresponding time evolution of magnetization $\phi$ (order parameter) after quenching. The magnetizations no longer evolve after reaching the steady-state values. The corresponding time step required to equilibrate the system (relaxation time) for given quenched fields are pointed through ($\otimes$) in Fig. \ref{timeConst_P} (a) and (b). The data displayed in this figure are calculated for system size $300^3$ under periodic boundary conditions with disorder strength $\sigma = 2.50$.}
\label{Mag_evolution}
\center
\end{figure}

Figures \ref{Mag_evolution}(a1) and (a2) graphically illustrates the phase ordering simulations where arrows indicate the direction of the single-step field quenched from the all-up or all-down spin configurations. After the quench, the system equilibrates through successive sets of spin-flips and finally reaches the steady-state value when it can no longer evolve due to the absence of thermal fluctuations. The time evolution of magnetization $\phi = \frac{1}{L^d}\sum_{i=1}^{L^d} s_i$ for decreasing and increasing quench are represented in Figs. \ref{Mag_evolution}(b1) and (b2). The total number of sets of spin-flips required to reach a steady-state configuration from the fully polarized state, termed as relaxation time constant, is plotted in Figs. \ref{timeConst_P} (a) and (b) as a function of quenched fields. The extraction procedure of relaxation time is detailed in the above algorithm (Sec. \ref{Phase-ordering}). 

\subsection{Dynamic hysteresis}
The dynamic hysteresis calculations involve a linear ramping of the field, ${H(t)=H_0 + R t}$ where $R$ is the rate of increasing or decreasing of the magnetic field across the transitions starting from an initial field $H_0 \to \pm \infty$. The magnetization of each step is calculated and presented as a function of the external field. The algorithm to compute the magnetization at each stage of increasing field is presented below.

\begin{enumerate}
\item We create a fully spin-polarized state by setting every site to  $s_i = -1$ for the initial magnetic field $H_0 \to - \infty$.
\item We increase the external field by $R$ in every time step i.e., $H(t)=H(t-1)+R$.
\item We recheck all the sites if the local field in Eq.~\ref{eqn:flip} changes the sign on any of the sites.
\item We flip the spins of the sites where the local field $E_i$ changes sign and then calculate the net magnetization corresponding to that external field.
\item Then, we proceed to the next time step by increasing the field by $R$ and repeating steps (2-5). We keep increasing the field until all the spins are flipped for large value of $H$, i.e., $s_i = 1$ for all $i$.
\end{enumerate}

The algorithm is not practical for the quasistatic simulations of ZTRFIM. Consequently, quasistatic loop, which corresponds to $R \rightarrow 0$ in our notation, is evaluated differently. We allow the system to equilibrate at each external field before increasing the field strength following Refs. \cite{Sethna_prl93, Sethna_prl95}. In quasistatic field change, the system's dynamic is unchanged; therefore, this process is often called adiabatic \cite{Sethna_prl93, Sanja_JSM21}. However, we cross-check that the linear ramping protocol with a prolonged ramp rate is in good agreement with the quasistatic protocol.

\begin{figure}[!h]
\center
\includegraphics[scale=0.3]{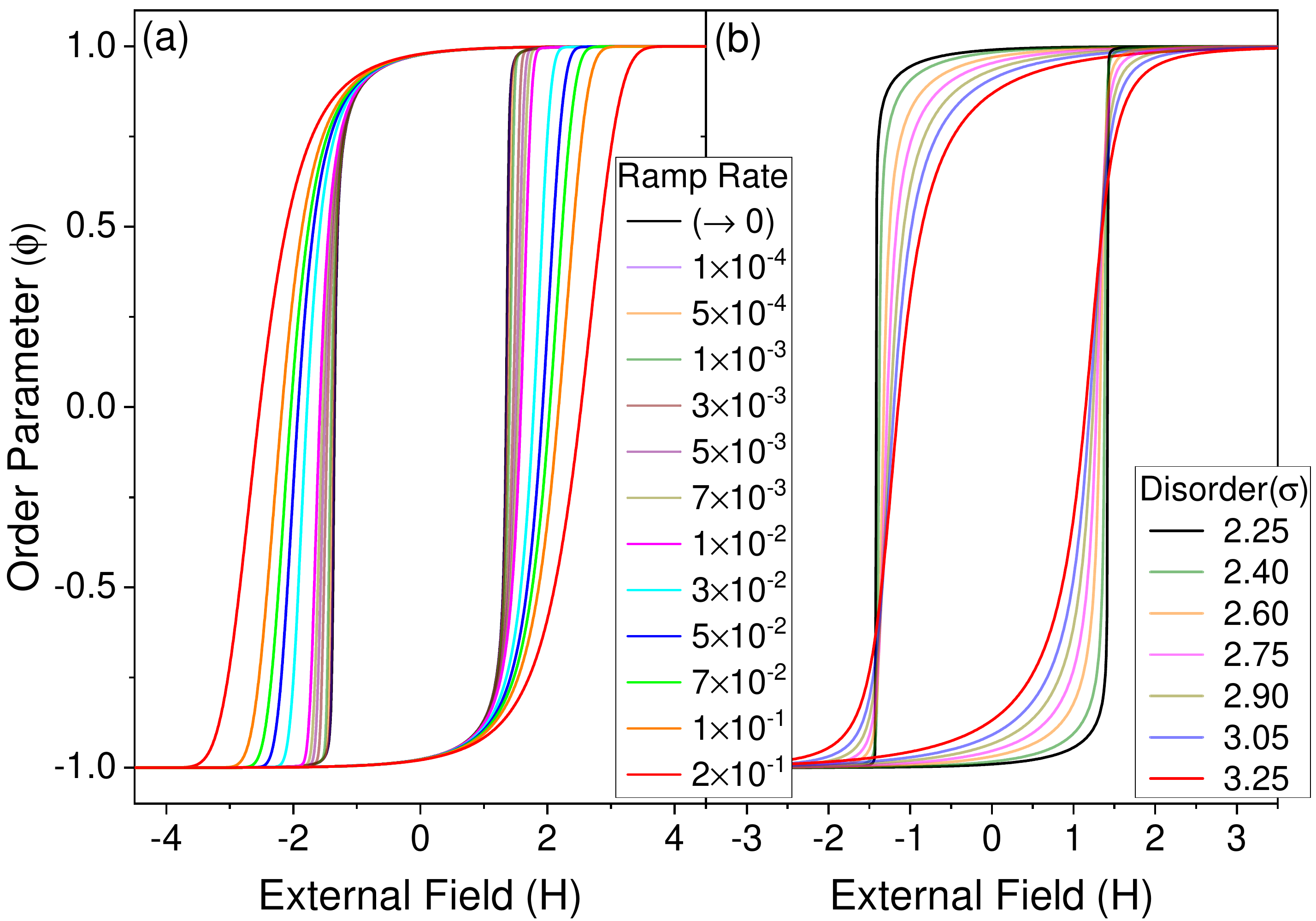}
\caption{(a) Magnetization $\phi$ as a function of external field $H$ for different ramp rates are computed using ZTRFIM simulations of $300^3$ system for disorder strength $\sigma = 2.50$. The quasistatic hysteresis curve is designated by ($\rightarrow$ 0). (b) The quasistatic hysteresis curve for different disorder strengths.}
\label{Dynamic_hysteresis}
\center
\end{figure}

Figure \ref{Dynamic_hysteresis}(a) shows hysteresis curves different ramping rate $R$. The area of the hysteresis curve increases systematically with the rate of change of the external field. The quasistatic hysteresis curve has also been extracted for different disorder strengths $\sigma$ [Fig. \ref{Dynamic_hysteresis}(b)]. As the disorder strength increases, the hysteresis width decreases, accompanied by a broader change of magnetization during switching. 

\subsection{Mean-field dynamic hysteresis}
We also performed a dynamic hysteresis simulation in the mean-field limit for the same model where the local field is controlled by the average magnetization. The local field is now defined as
\begin{equation}
E_i=J z \phi + h_i+H, 
\label{eqn:MFflip}
\end{equation}
where  $\phi = \frac{1}{L^d}\sum_{i=1}^{L^d} s_i$  is the average magnetization of the system, and $z$ is the number of nearest neighbors. Therefore, we carry out the mean-field dynamic hysteresis simulation by following the same algorithm discussed above using Eq. (\ref{eqn:MFflip}) instead of Eq. (\ref{eqn:flip}). The average magnetization of the spin-flip local field makes the system infinite range, which is equivalent to the mean-field approximation \cite{Sethna_prb96}.

\section{Results}

The simulated results introduce two separate phenomena emerging in phase ordering and dynamic hysteresis measurements.  The phase ordering dynamic captures the time scale of the relaxation and has been used to extract the critical disorder for a specific system, as discussed in the following section. We use this critical disorder as a boundary for the dynamic hysteresis measurements.

\subsection{Phase ordering time}

\begin{figure}[!h]
\center
\includegraphics[scale=0.31]{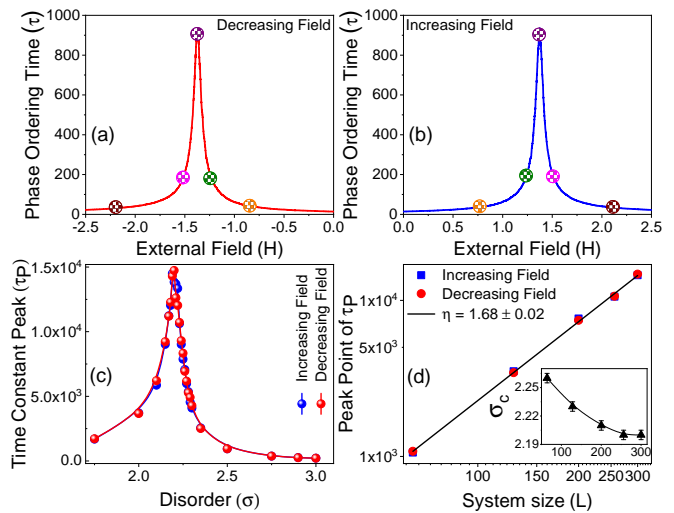}
\caption{Phase ordering relaxation time versus waiting fields for decreasing (a) and increasing (b) field-quenched. (c) The recorded peak-points of relaxation time $\tau$ as a function of  disorder strength ($\sigma$) for $300^3$ system. (d) The value at the maxima of the time-constant peaks $\tau_P$ corresponds to the classical-critical point of ZTRFIM and follows a finite size scaling: $(\tau_P)_{max} \propto L^{\eta}$, the exponent $\eta = 1.68\pm0.02$ in 3$d$. Inset shows the variation of critical disorder $\sigma_c$ with system size $L$.}
\label{timeConst_P}
\center
\end{figure}

The phase ordering time (relaxation time) of the quench-and-hold experiment is the total Monte Carlo time steps (total number of sets of spin-flips) to equilibrate the system onto the steady-state configuration. Figure \ref{timeConst_P} (a) and (b) show that the phase ordering time increases at the coercive field. The time constant peak at the field driven hysteretic transition points is the direct evidence of critical slowing down across the abrupt transformation \cite{CMBook_Lubensky}. Such slowing down in first-order transition can only occur when the system enters into the analytic regions of spinodal singularity \cite{Binder_pra84, Kundu_prl20, Zapperi_prl97, PTBook_Metastable}. This slowing down can also be observed in dynamic hysteresis measurements (see Sec. III B) in terms of the finite-time effect across the bifurcation points of hysteretic transition \cite{Alvaro_SciRep18}.

The time constant in the coercive field (the value at maxima of the phase ordering time) is plotted in Fig. \ref{timeConst_P}(c) as a function of disorder strength $\sigma$. There is a sharp increase of the time-constant peak $\tau_P$ around $\sigma = 2.20$, which corresponds to the classical-critical point of ZTRFIM \cite{Sethna_prl93, Sethna_prl95, Nandi_prl16}. At this point, the system shows a field-driven hysteretic first-order phase transition accompanied by a disorder-induced continuous transition. The extraction of the critical disorder has been a substantial task for the last three decades and it is still an ongoing exercise for different dimensions \cite{Mijatovic_physA21,  Mijatovic_pre19, Ahrens_prb11, Vives_pre99} and in infinite system size limits \cite{Fytas_prl13, Fytas_pre16}.  The rise in time-constant peak at the classical-critical point demonstrates that the phase-ordering dynamic is one such inventive technique for the extraction of the critical disorder. The values are in good qualitative agreement with the reported results (see Table \ref{table_vc}) in different dimensions \cite{Sethna_prl93, Sethna_prl95, Ahrens_prb11, Vives_pre99}. A little higher value of $\sigma_c$ has been observed as a reason for limited system-size calculation. Although we are not concerned about system size as the time constant peak follows a finite-size scaling; for example the scaling exponent $\eta = 1.68 \pm 0.02$ in 3$d$ [Fig. \ref{timeConst_P}(d)]. In the experiment, the avalanches are tricky to detect as the signal is often too low and smears outs in the bulk materials and in a higher driving rate \cite{Scheifele_pre13, Francisco_prl04}. In that case, the phase ordering dynamics can be applied easily \cite{Bar_prl18, Bar_prb21, Bray_AP02}.

\begin{figure}[!h]
\center
\includegraphics[scale=0.3]{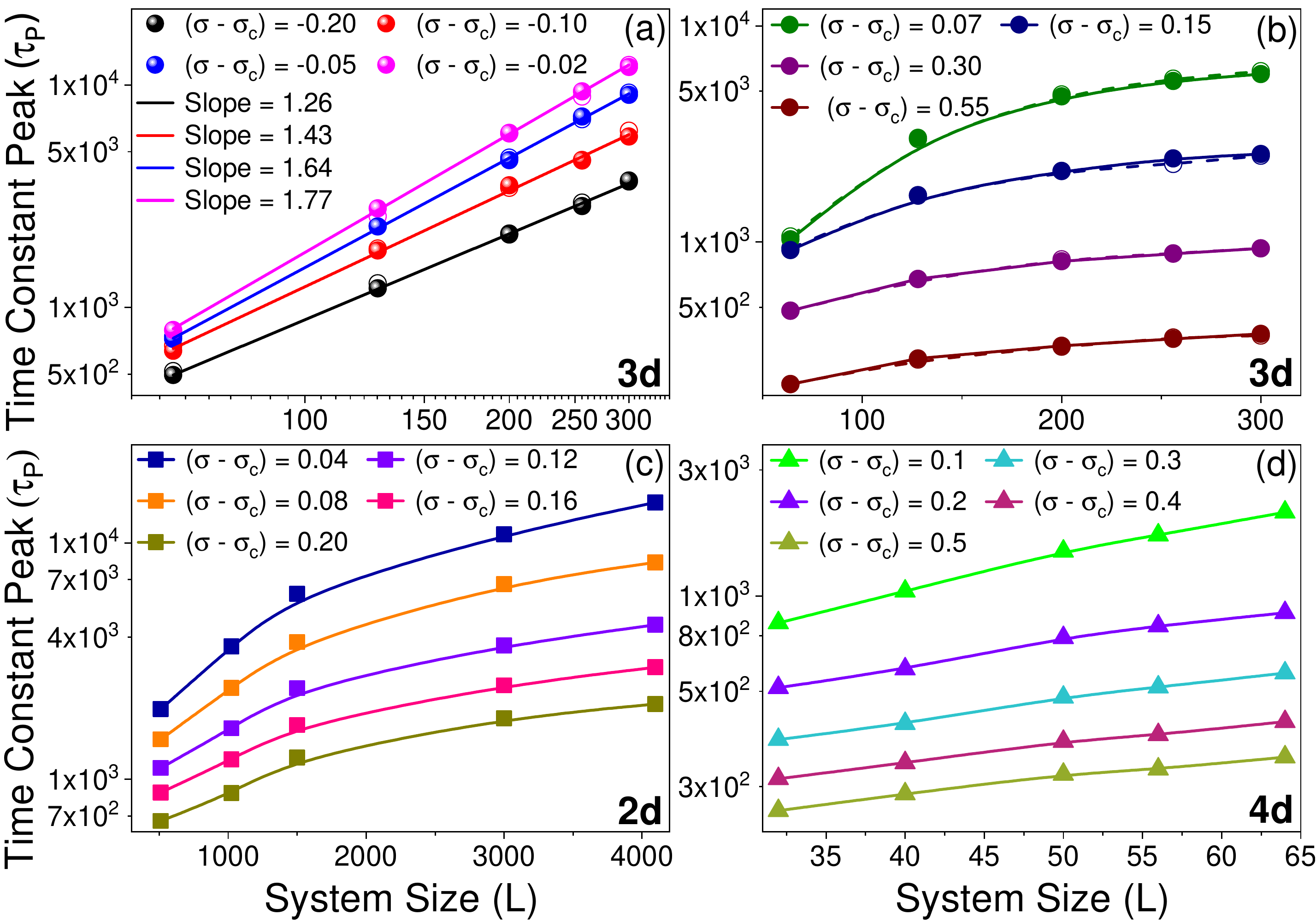}
\caption{The time-constant peaks $\tau_P$ below the classical-critical point follow power law finite-size effect (a), whereas it expresses a mild size effect above the critical point for $d = 3, 2$, and $4$ (b), (c), and (d). The solid and hollow symbols correspond to the increasing and decreasing of fields, and a little mismatch between them depends on how close the field reaches the transition points.}
\label{Finite_Size}
\center
\end{figure}

The finite-size effect of the time constant peak at the coercive fields behaves differently below and above the critical point (Fig. \ref{Finite_Size}). Below critical disorder, avalanche sizes are comparable to the system size; hence usual power-law scaling is expected [Fig. \ref{Finite_Size}(a)] \cite{Sanja_JSM21}. In contrast, the time-constant peak expands slowly with system size [Figs. \ref{Finite_Size}(b), (c), and (d)] for the high disorder systems suggesting mild critical slowing down due to the suppression of deep-rooted spinodal instability by quenched disorder \cite{ Bar_prb21, Bhowmik_pre19}. In other words, diverging correlation length (and susceptibility) becomes finite as it is bounded by the local fluctuations in disorder density in low dimension \cite{Imry_prb79}. Therefore, the spiky but nondiverging time constant peak at the coercive fields in Figs. \ref{timeConst_P}(a) and (b) is the effect of quenched disorder on the mean-field spinodal slowing down \cite{Nandi_prl16, Bar_prb21, Bhowmik_pre19, Scheifele_pre13}. The local fluctuations connected to the random disorder trigger the heterogeneous nucleation before reaching the spinodal point \cite{Cao_prb90, Imry_prb79, Fan_prb11, Scheifele_pre13, Wang_pre07}. Hence, such suppressed spinodal slowing down phenomena can be observed in low-dimensional systems.

\begin{table}
\caption{\label{table_vc} The critical disorder computed using phase ordering method for different dimensions. The value is compared with the reported data calculated using existing methods \cite{Sethna_prl95, Ahrens_prb11, Vives_pre99}.}
\begin{ruledtabular}
\begin{tabular}{l|l|l}
\textbf{Dimension} & \textbf{Phase ordering $\sigma_c$} & \textbf{Reported $\sigma_c$}\\
\hline
2$d$ & $0.69 $ & $0.64 $ \cite{Vives_pre99}\\
	&	&$0.54 $ \cite{Spasojevic_prl11}\\
\hline
3$d$ & $2.20 $ & $2.16 $ \cite{Sethna_prl95}\\
\hline
4$d$ & $4.12 $ & $4.18 $ \cite{Ahrens_prb11}\\
\hline
6$d$ & $7.75 $ & $7.78 $ \cite{Ahrens_prb11}\\
\end{tabular}
\end{ruledtabular}
\end{table}

\subsection{Dynamic hysteresis scaling}

We quantify the effect of disorder (for $\sigma > \sigma_c$) on the spinodal slowing down by determining the delay in the switching with the driving rate of the external field (i.e., finite-time measurements). During rapid measurements, the delay in the switching at the bifurcation point of hysteresis leads to a shift in the coercive field [Fig. \ref{Dynamic_hysteresis} (a)]. The shift in the coercive field in any finite-time measurements from the steady-state coercive field associated with infinite-time measurement  (or change in the hysteresis loop area from the steady-state loop area) follows a dynamic scaling with the rate of change of the external field ($R$) \cite{Lee_pre16, Yildiz_pre04, Wang_JCP11, He_prl93, Jiang_prb95, Pan_apl03, Liu_jap99, Kim_prb97, Jung_prl90, Bar_prb21, Bar_prl18, Kuang_ssc00, Wongdamnern_jap09, Wongdamnern_mcp10, wongdamnern_kem10, Yimnirun_apl06, Yimnirun_apl07, Zhang_ssc96, Pan_msc03, Rao_prb90, Rao_prb91, RaoComment_prl92, Zhong_FrontPhys17_Temp, Zhong_FrontPhys17_Field, Zhong_prl95, Zhong_prl05, Shukla_pre18}:
\begin{equation}
\Delta H_c (R)=|H_c(R)-H_c(0)|\propto R^\Upsilon.
\label{eq:Dynamic_Exp}
\end{equation}
where $H_c (R)$ is the coercive field for $R$, the rate of change of field, and $H_c(0)$ is the steady-state coercive field. The dynamic hysteresis scaling exponent $\Upsilon$ is essentially a ``finite-time scaling" analogous to finite-size scaling.  

In thermodynamic equilibrium, the correlation length diverges at the critical point of a continuous phase transition. Therefore, the system becomes scale-free and shows power-law scaling of various physical quantities. Due to the finite volume ($V = L^d$) of the system, the correlation length cannot diverge rather bounded by the system size $L$ \cite{PTBook_Goldenfeld, Henkel_prl98}. That eventually restricts the divergence of those physical quantities before the actual critical point ($\beta_{c} (V) < \beta_{c} (\infty); \beta_{c} = 1/(k_{B}T_{c})$). For example, the specific heat peak decreases with decreasing system size accompanied by a shift in transition point followed by a power-law scaling with system size $L$; $|T_c(\infty)-T_c(L)| = L^{-\lambda}$ where $\lambda$ indicate as a shift exponent \cite{Ferdinand_pr69}. One can draw a similar analogy of the shift in the transition point in the context of the metastable dynamics where the system is no longer in equilibrium \cite{Bar_prl18}. At the spinodal instability, the system shows a critical slowing down due to the divergence of characteristic time scale. Therefore, one would expect to observe a similar power-law scaling  [Eq.~(\ref{eq:Dynamic_Exp})] of the shift in transition points with the finite measurement times, i.e., the inverse of rates of change of driving field (or external driving parameter)\cite{Bar_prl18, Bar_prb21, Zhong_FrontPhys17_Temp, Zhong_FrontPhys17_Field, Zhong_prl05}. The exponent $\Upsilon$, parallel to the shift exponent when the system size $L$ is replaced by the rate of change of driving field $R$, is the quantitative measure of spinodal slowing down. In the mean-field calculation, the exponent $\Upsilon$ is always 2/3, which is the argument for the genuine spinodal transition (as it is a mean-field concept)\cite{Bar_prl18, Zhong_FrontPhys17_Temp, Zhong_FrontPhys17_Field, Zhong_prl05}. That can only be observed in long-range clean materials belonging to mean-field universality or in long-range disorder materials, provided disorder can be reorganized under training. However, in practice, quenched heterogeneities build up in numerous materials for various reasons such as underlying heterogeneous substrate, doping, pinning, friction, kinetic arrest, and many more. A complex interplay between quenched disorder and long-range force fields gives rise to disorder-associated athermal transition in those materials \cite{Book_Disorder}. As a result, non-mean-field exponents ($\Upsilon \neq 2/3$) have been observed in various materials [see Table \ref{table_exp}]. Such phenomena can be described through a nearest-neighbor Ising-like system in the zero-temperature environment \cite{Sethna_prl93, Sethna_prl95}. 

\begin{figure}[!h]
\center
\includegraphics[scale=0.27]{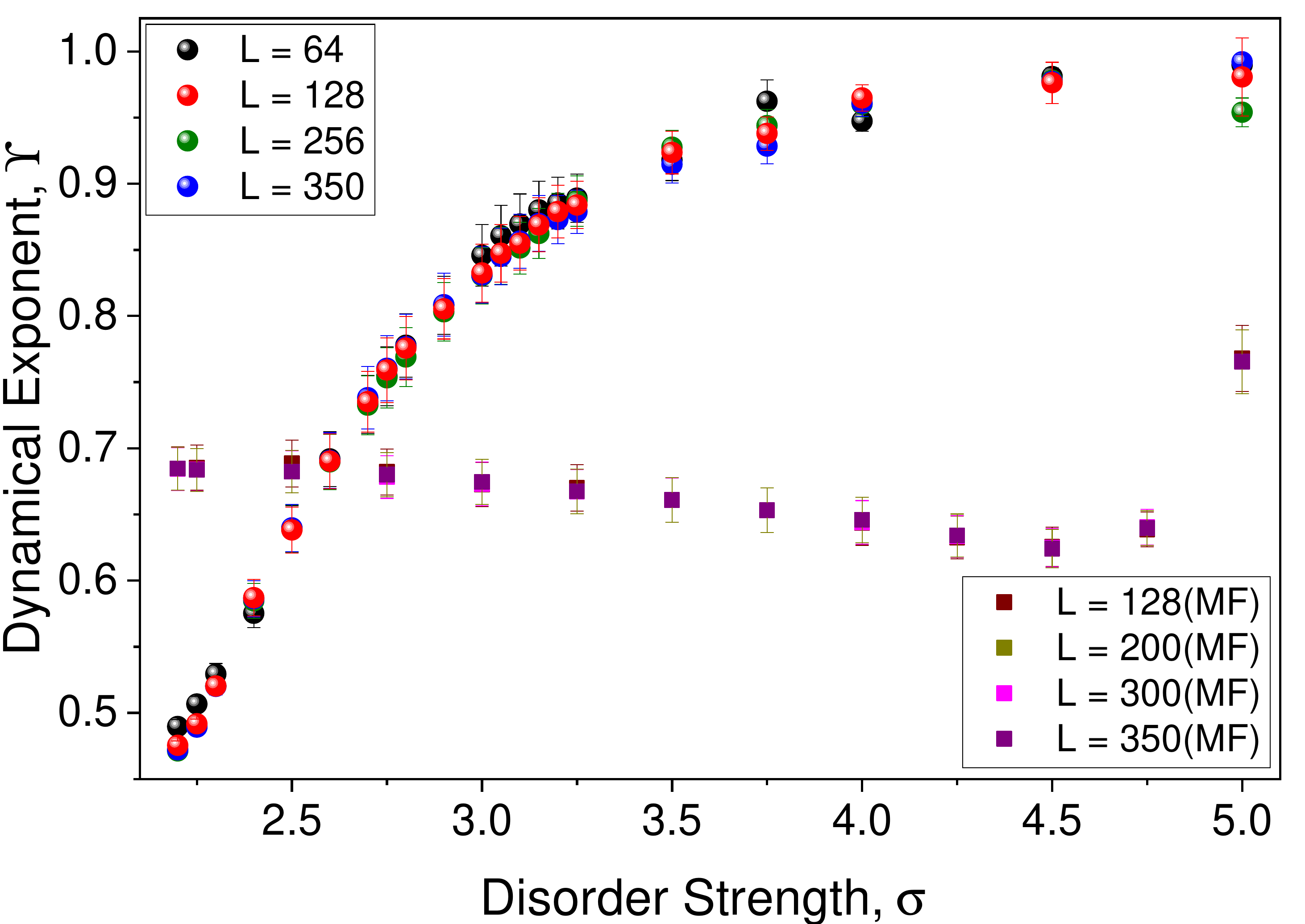}
\caption{Dynamical hysteresis scaling exponent ($\Upsilon$) extracted in the 3d-ZTRFIM simulations as a function of disorder strength ($\sigma$) for the following system sizes. ($\blacksquare$) represent the zero temperature mean-field dynamical exponent.}
\label{Exponent_3D}
\center
\end{figure}

The exponents $\Upsilon$, calculated in the 3d-ZTRFIM simulations, primarily increase with increasing disorder strength $\sigma$ and finally saturate to a value near $\Upsilon = 1$ (Fig. \ref{Exponent_3D}). While, in the mean-field limit, the value is consistent ($\Upsilon \approx 2/3$) within the uncertainty of the calculation except for $\sigma > 4.75$. The quasistatic transition in the mean-field model for disorder $\sigma > 4.75$ is no longer hysteretic, i.e. the switching is away from the saddle-node bifurcation point that violates the necessary conditions (hysteretic) of finite-time scaling [Eq.~(\ref{eq:Dynamic_Exp})]. Therefore, the sudden deviation of the scaling exponent from the mean-field value $\sigma = 5.0$ is insignificant in the context of this article. The error in exponent $\Upsilon$ increases during the crossover to the saturated value, and sometimes it does not follow a single exponent power-law scaling if the number of disorder average is low \cite{Bar_prb21}.  Most importantly, the dynamic hysteresis scaling exponent explicitly depends on the (diverging) time scale of the system as it is independent of system size. The finite-size effect has been canceling during steady-state subtraction [Eq.~(\ref{eq:Dynamic_Exp})]. 

\begin{figure}[!h]
\center
\includegraphics[scale=0.27]{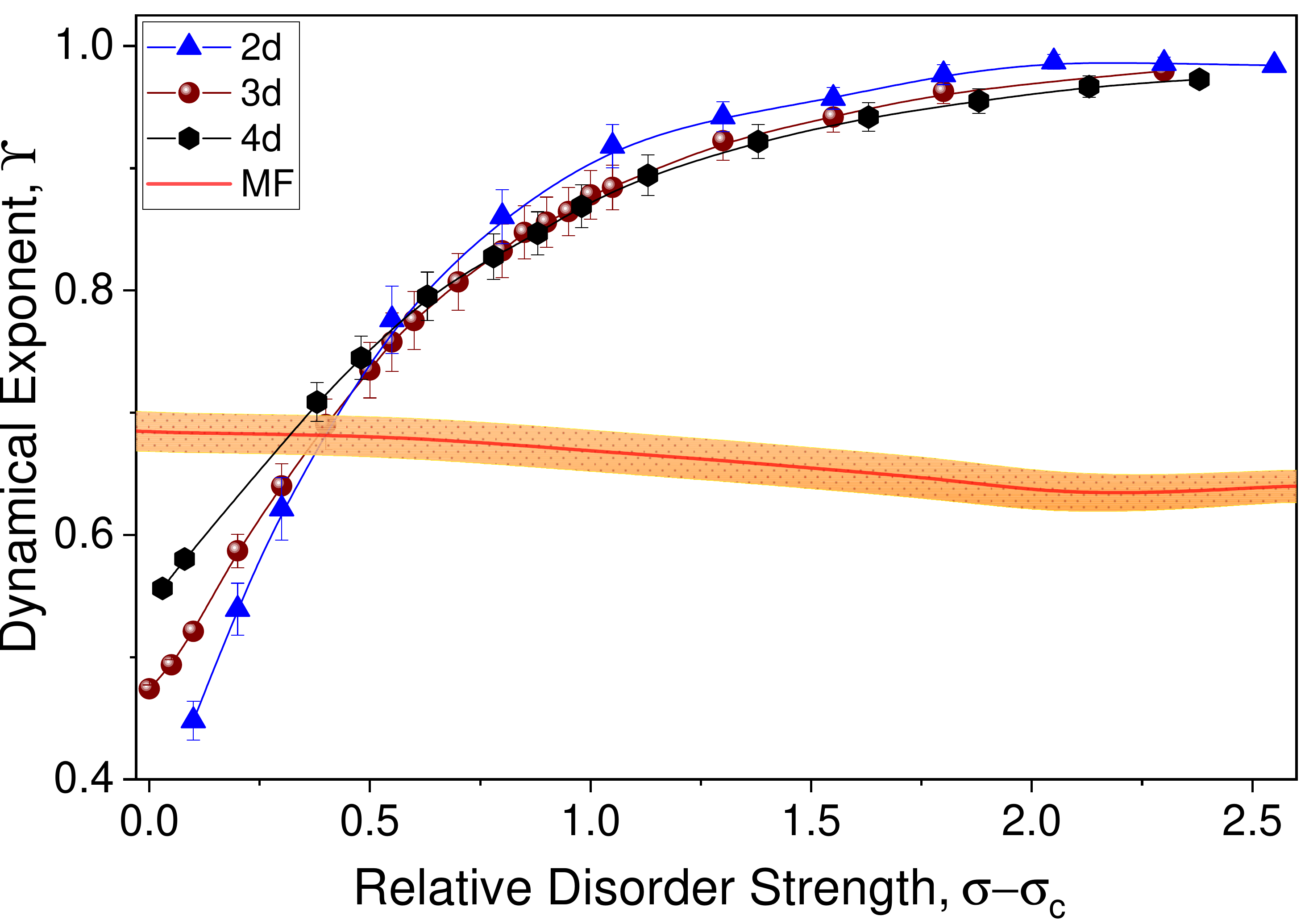}
\caption{Hysteresis scaling exponent ($\Upsilon$) versus disorder strength ($\sigma$) calculated at $d = 2, 3,$ and $4$ on hypercubic lattices. The shaded region represents the mean-field value, mimicking the infinite-dimensional calculation. The mean-field values for all disorder are independent of dimensions.}
\label{Exponent_allD}
\center
\end{figure}

To further investigate the spinodal slowing down in finite dimensions, we have computed  $\Upsilon$ versus $\sigma$, above $\sigma_c$, for different dimensions. With the increase in dimensionality, the exponent  $\Upsilon$ is slowly approaching towards mean-field value, which is fixed in all dimensions (Fig. \ref{Exponent_allD}). We have also observed the same trend even in six dimensions for small system sizes. Thus the normal upper critical dimension ($d_{up} = 6$) of ZTRFIM is not applicable for spinodal singularity \cite{Imry_prl75, Ahrens_prb11}. Therefore, we argue that genuine spinodal instability can be observed at very large (infinite) dimensions for such quenched-disorder systems, as suggested in the recent works \cite{Nandi_prl16, Berthier_prl20}.

\begin{table}
\caption{\label{table_exp} Experimentally reported dynamic hysteresis scaling exponent ($\Upsilon$) for different materials}
\begin{ruledtabular}
\begin{tabular}{l|l}
\textbf{System} & \textbf{Exponent $\Upsilon$}\\
\hline
Ferroelectric BaTiO3 single crystals \cite{Wongdamnern_jap09} & $0.195 \pm 0.016$\\
Ferroelectric BaTiO3 bulk ceramics \cite{wongdamnern_kem10} & $0.23 \pm 0.025$\\
Soft Pb(Ti, Zr)O$_3$ ferroelectric ceramic \cite{Yimnirun_apl06} & $0.25$\\
Hard Pb(Ti, Zr)O$_3$ ferroelectric ceramic \cite{Yimnirun_apl07} & $0.28 \pm 0.01$\\
Ultrathin Fe/Au ferromagnetic film \cite{He_prl93} & $0.31 \pm 0.05$\\
Ferroelectric Pb(Ti, Zr)O$_3$ thin film \cite{Liu_jap99} & $0.33$\\
Martensitic transition in Co (heating) \cite{Kuang_ssc00} & $0.39$\\
Polycrystalline BaTiO3 bulk ceramics  \cite{Wongdamnern_mcp10} & $0.39$\\
Antiferroelectric BPA mixed crystal \cite{Kim_prb97}& $0.40 \pm 0.04$\\
Martensitic transition in Co (cooling) \cite{Kuang_ssc00} & $0.49$\\
Structural transition in VO$_2$ \cite{Zhang_ssc96} & $0.51 \pm 0.09$\\
N-SmA transition in binary mixture (1:9) \cite{Yildiz_pre04} & $0.629 \pm 0.005$\\
Cold atomic system \cite{Lee_pre16} & $0.64 \pm 0.04$\\
Mott transition in V$_2$O$_3$ \cite{Bar_prl18} &  $0.66$\\
Ultrathin Co/Cu ferromagnetic film \cite{Jiang_prb95} & $0.66 \pm 0.03$\\
Ferroelectric SrBi$_2$Ta$_2$O$_9$ thin films \cite{Pan_apl03} & $0.66$\\
Switching of bistable laser \cite{Jung_prl90} &  $0.66$\\
N-SmA transition in binary mixture (4:6) \cite{Yildiz_pre04} & $0.672 \pm 0.008$\\
N-SmA transition in binary mixture (2:8) \cite{Yildiz_pre04} & $0.701 \pm 0.04$\\
N-SmA transition in binary mixture (3:7) \cite{Yildiz_pre04} & $0.766 \pm 0.05$\\
Martensitic transition in MnNiSn alloy \cite{Bar_prb21} &  $0.85 \pm 0.07$ \\
Glass transition of glycerol \cite{Wang_JCP11} & $0.88 \pm 0.09$\\
Austenite transition in MnNiSn alloy \cite{Bar_prb21} &  $0.93 \pm 0.13$ \\
Nickelate films with quenched disorder \cite{Prajapati_NJP22} &  $0.94 \pm 0.07$ \\
Nickelate films with quenched disorder \cite{Prajapati_NJP22} &  $0.98 \pm 0.04$ \\
PbTiO$_3$/polymer ferroelectric composites \cite{Pan_msc03} &  $1$ \\
Martensitic transition in FeMn alloy \cite{Zhang_ssc96} & $1$\\
\end{tabular}
\end{ruledtabular}
\end{table}

\section{Discussion}

It had long been considered that spinodals, an artifact of mean-field theories, cannot exist in low dimensions as any (thermal, disorder, or nonperturbative) fluctuations lead to overcoming the nucleation barrier before the spinodal can ever be reached \cite{PTBook_Metastable}. However, theories based on long-range interaction and coarse-grained Landau-Ginzburg formalism hint existence of spinodal instability in higher dimensions \cite{Binder_pra84, Ray_jsp90, Ray_jsp91}. Although, even above the upper critical dimension, it is difficult to observe mean-field-like spinodal in short-range force systems \cite{Nandi_prl16}. On the other hand, spinodal criticality can be seen in a low-dimensional system if the range of interaction exceeds some limiting value \cite{Gagliardi_pre21}. Therefore, theoretically, such instability is a matter of a competitive relationship between fluctuations, dimensionality, and the range of interaction of the system. In the long-ranged system, thermal fluctuation can be ignored \cite{Planes_prl01}, yet, the local fluctuation due to the disorder is an obstacle to the growth of susceptibility at the singular point \cite{Nandi_prl16}. In that case, the hidden instability is sometimes discernible only after numerous cycling (or training) through the transition point \citep{Bar_prl18}. The training may reorganize the disorder and naturally guide the system to approach the instability associated with self-organized criticality \cite{Francisco_prl07, Francisco_prb16}. A large class of materials (see Table \ref{table_exp}) shows the mean-field dynamic scaling exponent ($\Upsilon = 2/3$) connected to spinodal instability.  

Restructure of the disorder is not always achievable, precisely when the disorder is quenched. The local fluctuations linked with the disorder initiate a few heterogeneous nucleations on the pathways toward spinodal instability, where the growth is spontaneous due to the downhill nature of free energy. Finally, in the low dimension, the transformation takes place through a mixture of spinodal nucleation and classical nucleation and growth. Therefore, the criticality will remain hidden even above the upper critical dimension by the finite correlation length. That gives rise to a nonuniversal non-mean-field dynamic scaling exponent ($\Upsilon$) [Table \ref{table_exp}] accompanied by finite (not diverging) growth of phase ordering time in various experimental systems. The above arguments followed by ZTRFIM simulation capture nearly all the scaling exponent ($\Upsilon$) except ferroelectric switching. Due to large strains, the intrinsic domain-wall motion dominate  ferroelectric switching well below the curie temperature \cite{Liu_nat16}. The Coulomb forces are responsible for such switching that could make the system fundamentally different from the Ising-like \cite{Cohen_nat92}. However, the high disorder materials such as alloy, glass, and disordered nickelate by exhibiting higher exponents support the results obtained from ZTRFIM simulation (see Table \ref{table_exp} and Fig. \ref{Exponent_allD}). The other affirmation established from the upper limit of the exponents - $\Upsilon > 1$ has not been seen in any experiments to the best of our knowledge.

As disorder increases, emerging heterogeneous nucleating sites increase, leading to a decrease in the spinodal-nucleation process, and finally, for a sufficiently high disorder ($\sigma_{th} \approx \sigma_c+1.5$), the system is no longer critical. Above this threshold level, the nature of the supersaturated transition with the driving rate is independent of disorder strength. That could be identified as a distinct crossover from critical-like to a possible percolation-like transition  \cite{Shekhawat_prl13}.

Most importantly, dynamic scaling exponents ($\Upsilon$) approach toward mean-field values as we increase the dimensionality and is expected to approach the mean-field value only in an infinite dimension \cite{Berthier_prl20, Nandi_prl16} where the exponent is nearly independent of disorder strength ($\sigma$). 

\section{Conclusion}

The critical-like signatures such as diverging time-scale, diverging susceptibility, and observation of power-law scaling in an abrupt hysteresis transition in materials are directly linked with the spinodal instability \cite{Binder_pra84, Bar_prl18, Kundu_prl20, Zapperi_prl97}. The trademark of such instability can only be observed in a long-ranged interacting system where thermal fluctuation is irrelevant (athermal) such that the system is unable to hop the nucleation barrier of the parent phase \cite{Binder_pra84, Planes_prl01}. Based on the ZTRFIM simulation, we argue that the spinodal instability, even in an athermal system, gets hindered by the local fluctuations associated with quenched disorder \cite{Bar_prb21}. As the disorder increases in a finite-dimensional system, the associated fluctuation also increases that shield the instability accordingly. Finally, the transformation becomes conventional (non-critical) first-order above some threshold value of the disorder. Such hidden instability is gradually disclosed with the dimensionality of the system as nonperturbative local fluctuations reduces inversely with the dimension \cite{Nandi_prl16}. Therefore, non-mean-field critical behavior in abrupt hysteresis transitions is nothing but finite-dimensional vestiges of spinodal instability. This argument has recently been reported in glassy dynamics \cite{Berthier_prl20}. Here, we are presenting it from a general context that explain a large class of previously reported measurements in various materials being necessarily hysteretic.

\section{Acknowledgments}

It is a pleasure to thank Sanja Jani\'cevi\'c and Jordi Bar\'o i Urbea for critical comments and suggestions. T.B. thanks Gustau Catalan, Javier Rodr\'{i}guez-Viejo, and GTNaM members for discussion. T.B. acknowledges post-doctoral funding from ICN2 and Grant No. PID2019-108573GB-C21 funded by MCIN/AEI/10.13039/501100011033. The ICN2 is funded by the CERCA program/Generalitat de Catalunya. The ICN2 is supported by the Severo Ochoa program of MINECO (Grant No. SEV-2017-0706). A.B. acknowledge support from the Kreitman School of Advanced Graduate Studies and European Research Council (ERC) Grant Agreement No. 951541, ARO (W911NF-20-1-0013). The computations were performed on the BGU cluster.

\appendix
\section{Power-law fitting and error}
\label{App:Error}
\begin{figure}[!h]
\center
\includegraphics[scale=0.3]{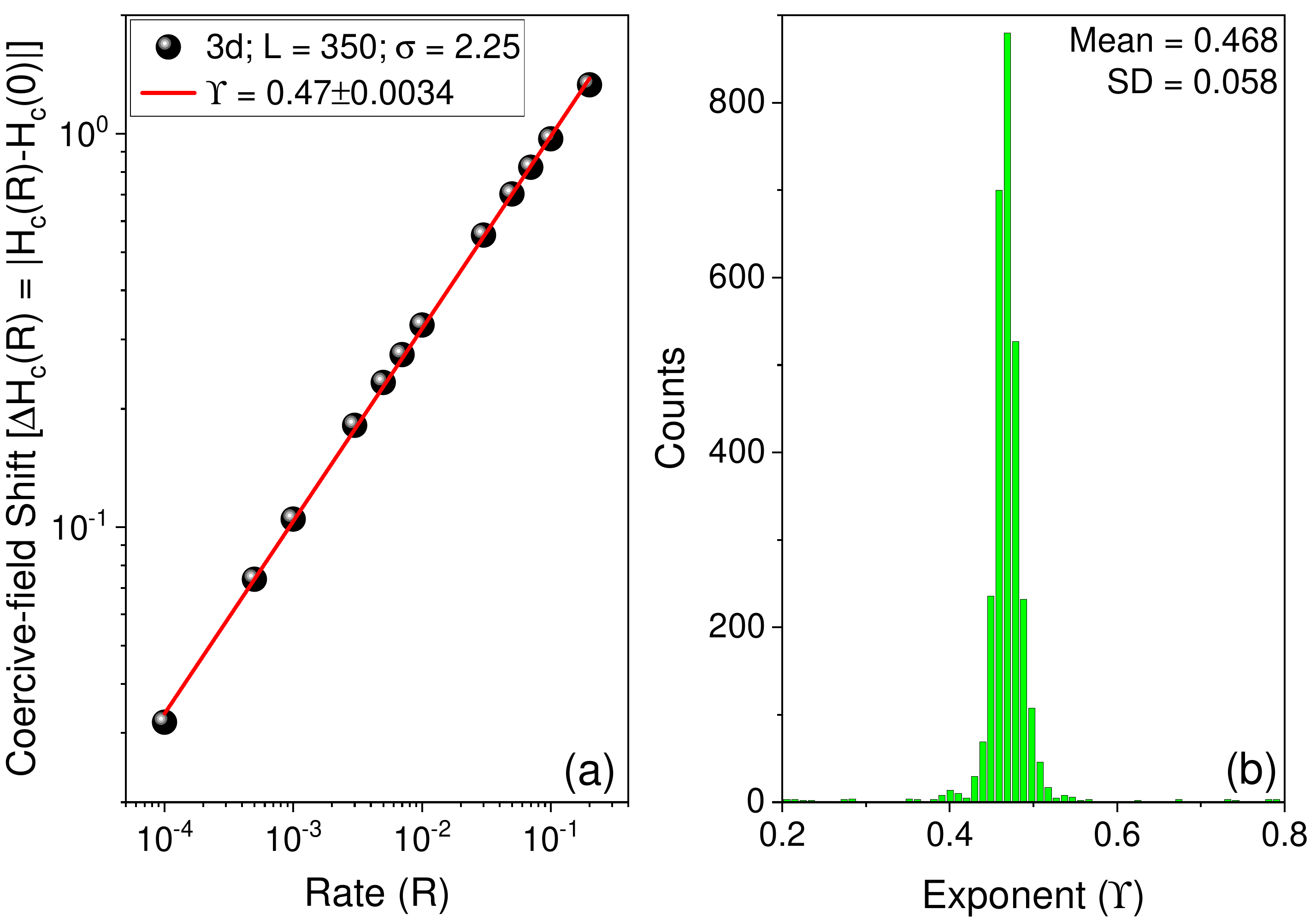}
\caption{(a) Log-log plot of shift in coercive field from the steady-state coercive field with rate of change of field ($\bullet$). (---) represent the power laws fitting with exponent $\Upsilon = 0.47 \pm 0.0034$. (b) Histogram of fitting exponent $\Upsilon$. The exponents were evaluated by choosing four independent points out of the whole data set. The displayed analyses have been done on data set of increasing field for $L^d = 350^3$ and $\sigma = 2.25$.}
\label{Histogram}
\center
\end{figure}

The dynamical shifts in coercive fields from the steady-state coercive field follow a scaling with the rate of change of external field $R$. In the scaling equation [Eq. (\ref{eq:Dynamic_Exp})], there is only one fitting parameter, i.e., the exponent $\Upsilon$. The exponent has been extracted by fitting a straight line in the log-log graph, where the slope of the straight line specifies the value of $\Upsilon$ [Fig. \ref{Histogram}(a)]. 

In the log-log graph, the fitting is dominated by the lower rate values and the steady-state coercive field $H_c(0)$. The inaccuracy in $H_c$ for lower $R$ may lead to a large error in the exponent value; specifically, a small error in $H_c(0)$ could ruin the fittings. We cross-check each fitting exponent using another rational fitting tool where each data point plays an equal role in extracting the exponent.

{\it Statistical distributions of nonlinear fitting:} In this technique, we pick up four data points from the complete set of data corresponding to different rates and calculate the exponents for all possible combinations. Using those exponents, we calculate the steady-state coercive field respectively. We consider only those exponents that lie between the numerical uncertainty of the coercive field corresponding to the relative variance of magnetization. The distribution of accepted exponents obeys a normal distribution. The distributions' mean and standard deviation can be considered the effective exponent and corresponding error. The details of the technique are following.

Let us assume ${H_c}_i$ and ${H_c}_j$ are the coercive field for $i$-th and $j$-th rate of change of field. From Eq. (\ref{eq:Dynamic_Exp}), the shift in coercive field from the steady-state coercive field $H_c(0)$ can be written as
\begin{equation} \label{error1}
{H_c}_{i} = {H_c}(0) + aR_{i}^{\Upsilon}, \hspace{10mm} {H_c}_{j} = {H_c}(0) + aR_{j}^{\Upsilon}.  
\end{equation}
The sign of the constant $a$ depends upon the decreasing and increasing field. 
The influence of $H_c(0)$ for the extraction of the exponent can be abolished by subtracting the above two equations,
\begin{equation} \label{error2}
({H_c}_i - {H_c}_j) = a(R_{i}^{\Upsilon} - R_{j}^{\Upsilon}).
\end{equation}
If N is the total number of field rate we have $^NC_2$ (say $N1$) similar equations. We eliminate the constant $a$ by dividing any two such equations (for example, $(i, j)$ and $(k, l)$ pairs), i.e.,
\begin{equation} \label{error3}
\frac{({H_c}_i - {H_c}_j)}{({H_c}_k - {H_c}_l)} = \frac{(R_{i}^{\Upsilon} - R_{j}^{\Upsilon})}{(R_{k}^{\Upsilon} - R_{l}^{\Upsilon})}.
\end{equation}
Here $(i, j) \neq (k, l)$ ; but we count combinations such as $i = k $ if $j \neq l$ and vice versa. Therefore, one can pick up two pairs in $^{N1}C_2$ possible ways.  Numerical solutions of $^{N1}C_2$ transcendental equations supply $^{N1}C_2$ numbers of $\Upsilon$ that are free from all kinds of technical domination. By plotting the distribution of $^{N1}C_2$ number of $\Upsilon$, one can examine whether this data set follows a power-law scaling at all. For example, the data set does not follow a scaling law if one gets any other distribution rather than a sharp(within the acceptable error) normal distribution.

However, for each data pair $(i, j)$ there is one $H_c(0)$. 

\begin{equation}
H_c(0)^{\{i,j\}} = \frac{{H_c}_i - (\frac{R_i}{R_j})^\Upsilon {H_c}_j}{1-(\frac{R_i}{R_j})^\Upsilon},
\end{equation}
\begin{equation}
H_c(0)^{\{k, l\}} = \frac{{H_c}_k - (\frac{R_k}{R_l})^\Upsilon {H_c}_l}{1-(\frac{R_k}{R_l})^\Upsilon}
\end{equation}

There is no limitation on the value of $H_c(0)$ that is not justifiable for the monotonic increasing function of Eq. (\ref{error1}). The numerical errors of the two points in a random pair may yield some unacceptable $H_c(0)$ along with an incorrect $\Upsilon$. To draw an accurate distribution of $\Upsilon$, we neglected some values of $\Upsilon$ for which the inferred $H_c(0)$ lying outside the uncertainty of coercive field corresponds to the relative variance of magnetization [$H_c(0) \pm \delta {H_c(0)}$].

The mean of the distribution [Fig. \ref{Histogram}(b)], $\Upsilon_{mean} = 0.468$, is in good agreement with the straight-line fitting exponent [Fig. \ref{Histogram}(a)]. The standard deviation of the distribution is larger than the least square-fitting error. In the non-linear fitting method, a small numerical inaccuracy in coercive fields ($H_c(R) \pm \delta {H_c(R)}$) for any rate ($R$) magnifies the error of the exponent, which is over-estimated. Therefore, we have reported the least squares-fitting error in Figs. \ref{Exponent_3D} and \ref{Exponent_allD}. Note that the goodness of the fitting has been observed in the statistical distribution for all the data sets.

\bibliography{ZTRFIMRef.bib}

\begin{thebibliography}{82}%
\makeatletter
\providecommand \@ifxundefined [1]{%
 \@ifx{#1\undefined}
}%
\providecommand \@ifnum [1]{%
 \ifnum #1\expandafter \@firstoftwo
 \else \expandafter \@secondoftwo
 \fi
}%
\providecommand \@ifx [1]{%
 \ifx #1\expandafter \@firstoftwo
 \else \expandafter \@secondoftwo
 \fi
}%
\providecommand \natexlab [1]{#1}%
\providecommand \enquote  [1]{``#1''}%
\providecommand \bibnamefont  [1]{#1}%
\providecommand \bibfnamefont [1]{#1}%
\providecommand \citenamefont [1]{#1}%
\providecommand \href@noop [0]{\@secondoftwo}%
\providecommand \href [0]{\begingroup \@sanitize@url \@href}%
\providecommand \@href[1]{\@@startlink{#1}\@@href}%
\providecommand \@@href[1]{\endgroup#1\@@endlink}%
\providecommand \@sanitize@url [0]{\catcode `\\12\catcode `\$12\catcode
  `\&12\catcode `\#12\catcode `\^12\catcode `\_12\catcode `\%12\relax}%
\providecommand \@@startlink[1]{}%
\providecommand \@@endlink[0]{}%
\providecommand \url  [0]{\begingroup\@sanitize@url \@url }%
\providecommand \@url [1]{\endgroup\@href {#1}{\urlprefix }}%
\providecommand \urlprefix  [0]{URL }%
\providecommand \Eprint [0]{\href }%
\providecommand \doibase [0]{http://dx.doi.org/}%
\providecommand \selectlanguage [0]{\@gobble}%
\providecommand \bibinfo  [0]{\@secondoftwo}%
\providecommand \bibfield  [0]{\@secondoftwo}%
\providecommand \translation [1]{[#1]}%
\providecommand \BibitemOpen [0]{}%
\providecommand \bibitemStop [0]{}%
\providecommand \bibitemNoStop [0]{.\EOS\space}%
\providecommand \EOS [0]{\spacefactor3000\relax}%
\providecommand \BibitemShut  [1]{\csname bibitem#1\endcsname}%
\let\auto@bib@innerbib\@empty
\bibitem [{\citenamefont {Post}\ \emph {et~al.}(2018)\citenamefont {Post},
  \citenamefont {McLeod}, \citenamefont {Hepting}, \citenamefont {Bluschke},
  \citenamefont {Wang}, \citenamefont {Cristiani}, \citenamefont {Logvenov},
  \citenamefont {Charnukha}, \citenamefont {Ni}, \citenamefont {Radhakrishnan},
  \citenamefont {Minola}, \citenamefont {Pasupathy}, \citenamefont {Boris},
  \citenamefont {Benckiser}, \citenamefont {Dahmen}, \citenamefont {Carlson},
  \citenamefont {Keimer},\ and\ \citenamefont {Basov}}]{Basov_Nat18}%
  \BibitemOpen
  \bibfield  {author} {\bibinfo {author} {\bibfnamefont {K.~W.}\ \bibnamefont
  {Post}}, \bibinfo {author} {\bibfnamefont {A.~S.}\ \bibnamefont {McLeod}},
  \bibinfo {author} {\bibfnamefont {M.}~\bibnamefont {Hepting}}, \bibinfo
  {author} {\bibfnamefont {M.}~\bibnamefont {Bluschke}}, \bibinfo {author}
  {\bibfnamefont {Y.}~\bibnamefont {Wang}}, \bibinfo {author} {\bibfnamefont
  {G.}~\bibnamefont {Cristiani}}, \bibinfo {author} {\bibfnamefont
  {G.}~\bibnamefont {Logvenov}}, \bibinfo {author} {\bibfnamefont
  {A.}~\bibnamefont {Charnukha}}, \bibinfo {author} {\bibfnamefont {G.~X.}\
  \bibnamefont {Ni}}, \bibinfo {author} {\bibfnamefont {P.}~\bibnamefont
  {Radhakrishnan}}, \bibinfo {author} {\bibfnamefont {M.}~\bibnamefont
  {Minola}}, \bibinfo {author} {\bibfnamefont {A.}~\bibnamefont {Pasupathy}},
  \bibinfo {author} {\bibfnamefont {A.~V.}\ \bibnamefont {Boris}}, \bibinfo
  {author} {\bibfnamefont {E.}~\bibnamefont {Benckiser}}, \bibinfo {author}
  {\bibfnamefont {K.~A.}\ \bibnamefont {Dahmen}}, \bibinfo {author}
  {\bibfnamefont {E.~W.}\ \bibnamefont {Carlson}}, \bibinfo {author}
  {\bibfnamefont {B.}~\bibnamefont {Keimer}}, \ and\ \bibinfo {author}
  {\bibfnamefont {D.~N.}\ \bibnamefont {Basov}},\ }\href {\doibase
  10.1038/s41567-018-0201-1} {\bibfield  {journal} {\bibinfo  {journal} {Nat.
  Phys.}\ }\textbf {\bibinfo {volume} {14}},\ \bibinfo {pages} {1056} (\bibinfo
  {year} {2018})}\BibitemShut {NoStop}%
\bibitem [{\citenamefont {McLeod}\ \emph {et~al.}(2017)\citenamefont {McLeod},
  \citenamefont {Van~Heumen}, \citenamefont {Ramirez}, \citenamefont {Wang},
  \citenamefont {Saerbeck}, \citenamefont {Guenon}, \citenamefont {Goldflam},
  \citenamefont {Anderegg}, \citenamefont {Kelly}, \citenamefont {Mueller},
  \citenamefont {Liu}, \citenamefont {Schuller},\ and\ \citenamefont
  {Basov}}]{Basov_Nat17}%
  \BibitemOpen
  \bibfield  {author} {\bibinfo {author} {\bibfnamefont {A.~S.}\ \bibnamefont
  {McLeod}}, \bibinfo {author} {\bibfnamefont {E.}~\bibnamefont {Van~Heumen}},
  \bibinfo {author} {\bibfnamefont {J.~G.}\ \bibnamefont {Ramirez}}, \bibinfo
  {author} {\bibfnamefont {S.}~\bibnamefont {Wang}}, \bibinfo {author}
  {\bibfnamefont {T.}~\bibnamefont {Saerbeck}}, \bibinfo {author}
  {\bibfnamefont {S.}~\bibnamefont {Guenon}}, \bibinfo {author} {\bibfnamefont
  {M.}~\bibnamefont {Goldflam}}, \bibinfo {author} {\bibfnamefont
  {L.}~\bibnamefont {Anderegg}}, \bibinfo {author} {\bibfnamefont
  {P.}~\bibnamefont {Kelly}}, \bibinfo {author} {\bibfnamefont
  {A.}~\bibnamefont {Mueller}}, \bibinfo {author} {\bibfnamefont {M.~K.}\
  \bibnamefont {Liu}}, \bibinfo {author} {\bibfnamefont {I.~K.}\ \bibnamefont
  {Schuller}}, \ and\ \bibinfo {author} {\bibfnamefont {D.~N.}\ \bibnamefont
  {Basov}},\ }\href {\doibase 10.1038/nphys3882} {\bibfield  {journal}
  {\bibinfo  {journal} {Nat. Phys.}\ }\textbf {\bibinfo {volume} {13}},\
  \bibinfo {pages} {80} (\bibinfo {year} {2017})}\BibitemShut {NoStop}%
\bibitem [{\citenamefont {Bar}\ \emph {et~al.}(2018)\citenamefont {Bar},
  \citenamefont {Choudhary}, \citenamefont {Ashraf}, \citenamefont {Sujith},
  \citenamefont {Puri}, \citenamefont {Raj},\ and\ \citenamefont
  {Bansal}}]{Bar_prl18}%
  \BibitemOpen
  \bibfield  {author} {\bibinfo {author} {\bibfnamefont {T.}~\bibnamefont
  {Bar}}, \bibinfo {author} {\bibfnamefont {S.~K.}\ \bibnamefont {Choudhary}},
  \bibinfo {author} {\bibfnamefont {M.~A.}\ \bibnamefont {Ashraf}}, \bibinfo
  {author} {\bibfnamefont {K.~S.}\ \bibnamefont {Sujith}}, \bibinfo {author}
  {\bibfnamefont {S.}~\bibnamefont {Puri}}, \bibinfo {author} {\bibfnamefont
  {S.}~\bibnamefont {Raj}}, \ and\ \bibinfo {author} {\bibfnamefont
  {B.}~\bibnamefont {Bansal}},\ }\href {\doibase
  10.1103/PhysRevLett.121.045701} {\bibfield  {journal} {\bibinfo  {journal}
  {Phys. Rev. Lett.}\ }\textbf {\bibinfo {volume} {121}},\ \bibinfo {pages}
  {045701} (\bibinfo {year} {2018})}\BibitemShut {NoStop}%
\bibitem [{\citenamefont {Kundu}\ \emph {et~al.}(2020)\citenamefont {Kundu},
  \citenamefont {Bar}, \citenamefont {Nayak},\ and\ \citenamefont
  {Bansal}}]{Kundu_prl20}%
  \BibitemOpen
  \bibfield  {author} {\bibinfo {author} {\bibfnamefont {S.}~\bibnamefont
  {Kundu}}, \bibinfo {author} {\bibfnamefont {T.}~\bibnamefont {Bar}}, \bibinfo
  {author} {\bibfnamefont {R.~K.}\ \bibnamefont {Nayak}}, \ and\ \bibinfo
  {author} {\bibfnamefont {B.}~\bibnamefont {Bansal}},\ }\href {\doibase
  10.1103/PhysRevLett.124.095703} {\bibfield  {journal} {\bibinfo  {journal}
  {Phys. Rev. Lett.}\ }\textbf {\bibinfo {volume} {124}},\ \bibinfo {pages}
  {095703} (\bibinfo {year} {2020})}\BibitemShut {NoStop}%
\bibitem [{\citenamefont {Chandni}\ \emph {et~al.}(2009)\citenamefont
  {Chandni}, \citenamefont {Ghosh}, \citenamefont {Vijaya},\ and\ \citenamefont
  {Mohan}}]{Chandni_prl09}%
  \BibitemOpen
  \bibfield  {author} {\bibinfo {author} {\bibfnamefont {U.}~\bibnamefont
  {Chandni}}, \bibinfo {author} {\bibfnamefont {A.}~\bibnamefont {Ghosh}},
  \bibinfo {author} {\bibfnamefont {H.~S.}\ \bibnamefont {Vijaya}}, \ and\
  \bibinfo {author} {\bibfnamefont {S.}~\bibnamefont {Mohan}},\ }\href
  {\doibase 10.1103/PhysRevLett.102.025701} {\bibfield  {journal} {\bibinfo
  {journal} {Phys. Rev. Lett.}\ }\textbf {\bibinfo {volume} {102}},\ \bibinfo
  {pages} {025701} (\bibinfo {year} {2009})}\BibitemShut {NoStop}%
\bibitem [{\citenamefont {Bar}\ \emph {et~al.}(2021)\citenamefont {Bar},
  \citenamefont {Ghosh},\ and\ \citenamefont {Banerjee}}]{Bar_prb21}%
  \BibitemOpen
  \bibfield  {author} {\bibinfo {author} {\bibfnamefont {T.}~\bibnamefont
  {Bar}}, \bibinfo {author} {\bibfnamefont {A.}~\bibnamefont {Ghosh}}, \ and\
  \bibinfo {author} {\bibfnamefont {A.}~\bibnamefont {Banerjee}},\ }\href
  {\doibase 10.1103/PhysRevB.104.144102} {\bibfield  {journal} {\bibinfo
  {journal} {Phys. Rev. B}\ }\textbf {\bibinfo {volume} {104}},\ \bibinfo
  {pages} {144102} (\bibinfo {year} {2021})}\BibitemShut {NoStop}%
\bibitem [{\citenamefont {Keim}\ \emph {et~al.}(2019)\citenamefont {Keim},
  \citenamefont {Paulsen}, \citenamefont {Zeravcic}, \citenamefont {Sastry},\
  and\ \citenamefont {Nagel}}]{Keim_rmp19}%
  \BibitemOpen
  \bibfield  {author} {\bibinfo {author} {\bibfnamefont {N.~C.}\ \bibnamefont
  {Keim}}, \bibinfo {author} {\bibfnamefont {J.~D.}\ \bibnamefont {Paulsen}},
  \bibinfo {author} {\bibfnamefont {Z.}~\bibnamefont {Zeravcic}}, \bibinfo
  {author} {\bibfnamefont {S.}~\bibnamefont {Sastry}}, \ and\ \bibinfo {author}
  {\bibfnamefont {S.~R.}\ \bibnamefont {Nagel}},\ }\href {\doibase
  10.1103/RevModPhys.91.035002} {\bibfield  {journal} {\bibinfo  {journal}
  {Rev. Mod. Phys.}\ }\textbf {\bibinfo {volume} {91}},\ \bibinfo {pages}
  {035002} (\bibinfo {year} {2019})}\BibitemShut {NoStop}%
\bibitem [{\citenamefont {T\'oth}\ \emph {et~al.}(2014)\citenamefont {T\'oth},
  \citenamefont {Szab\'o}, \citenamefont {Dar\'oczi},\ and\ \citenamefont
  {Beke}}]{Toth_prb14}%
  \BibitemOpen
  \bibfield  {author} {\bibinfo {author} {\bibfnamefont {L.}~\bibnamefont
  {T\'oth}}, \bibinfo {author} {\bibfnamefont {S.}~\bibnamefont {Szab\'o}},
  \bibinfo {author} {\bibfnamefont {L.}~\bibnamefont {Dar\'oczi}}, \ and\
  \bibinfo {author} {\bibfnamefont {D.~L.}\ \bibnamefont {Beke}},\ }\href
  {\doibase 10.1103/PhysRevB.90.224103} {\bibfield  {journal} {\bibinfo
  {journal} {Phys. Rev. B}\ }\textbf {\bibinfo {volume} {90}},\ \bibinfo
  {pages} {224103} (\bibinfo {year} {2014})}\BibitemShut {NoStop}%
\bibitem [{\citenamefont {Gallardo}\ \emph {et~al.}(2010)\citenamefont
  {Gallardo}, \citenamefont {Manchado}, \citenamefont {Romero}, \citenamefont
  {del Cerro}, \citenamefont {Salje}, \citenamefont {Planes}, \citenamefont
  {Vives}, \citenamefont {Romero},\ and\ \citenamefont
  {Stipcich}}]{Gallardo_prb10}%
  \BibitemOpen
  \bibfield  {author} {\bibinfo {author} {\bibfnamefont {M.~C.}\ \bibnamefont
  {Gallardo}}, \bibinfo {author} {\bibfnamefont {J.}~\bibnamefont {Manchado}},
  \bibinfo {author} {\bibfnamefont {F.~J.}\ \bibnamefont {Romero}}, \bibinfo
  {author} {\bibfnamefont {J.}~\bibnamefont {del Cerro}}, \bibinfo {author}
  {\bibfnamefont {E.~K.~H.}\ \bibnamefont {Salje}}, \bibinfo {author}
  {\bibfnamefont {A.}~\bibnamefont {Planes}}, \bibinfo {author} {\bibfnamefont
  {E.}~\bibnamefont {Vives}}, \bibinfo {author} {\bibfnamefont
  {R.}~\bibnamefont {Romero}}, \ and\ \bibinfo {author} {\bibfnamefont
  {M.}~\bibnamefont {Stipcich}},\ }\href {\doibase 10.1103/PhysRevB.81.174102}
  {\bibfield  {journal} {\bibinfo  {journal} {Phys. Rev. B}\ }\textbf {\bibinfo
  {volume} {81}},\ \bibinfo {pages} {174102} (\bibinfo {year}
  {2010})}\BibitemShut {NoStop}%
\bibitem [{\citenamefont {Kakeshita}\ \emph {et~al.}(2011)\citenamefont
  {Kakeshita}, \citenamefont {Fukuda}, \citenamefont {Saxena},\ and\
  \citenamefont {Planes}}]{Book_Disorder}%
  \BibitemOpen
  \bibfield  {author} {\bibinfo {author} {\bibfnamefont {T.}~\bibnamefont
  {Kakeshita}}, \bibinfo {author} {\bibfnamefont {T.}~\bibnamefont {Fukuda}},
  \bibinfo {author} {\bibfnamefont {A.}~\bibnamefont {Saxena}}, \ and\ \bibinfo
  {author} {\bibfnamefont {A.}~\bibnamefont {Planes}},\ }\href
  {https://books.google.co.in/books?id=KwKTXX5GxaQC} {\emph {\bibinfo {title}
  {Disorder and Strain-Induced Complexity in Functional Materials}}},\ Springer
  Series in Materials Science\ (\bibinfo  {publisher} {Springer Berlin
  Heidelberg},\ \bibinfo {year} {2011})\BibitemShut {NoStop}%
\bibitem [{\citenamefont {Parisi}\ \emph {et~al.}(2017)\citenamefont {Parisi},
  \citenamefont {Procaccia}, \citenamefont {Rainone},\ and\ \citenamefont
  {Singh}}]{Parisi_PNAS17}%
  \BibitemOpen
  \bibfield  {author} {\bibinfo {author} {\bibfnamefont {G.}~\bibnamefont
  {Parisi}}, \bibinfo {author} {\bibfnamefont {I.}~\bibnamefont {Procaccia}},
  \bibinfo {author} {\bibfnamefont {C.}~\bibnamefont {Rainone}}, \ and\
  \bibinfo {author} {\bibfnamefont {M.}~\bibnamefont {Singh}},\ }\href
  {\doibase 10.1073/pnas.1700075114} {\bibfield  {journal} {\bibinfo  {journal}
  {Proc. Natl. Acad. Sci. U.S.A.}\ }\textbf {\bibinfo {volume} {114}},\
  \bibinfo {pages} {5577} (\bibinfo {year} {2017})}\BibitemShut {NoStop}%
\bibitem [{\citenamefont {Ozawa}\ \emph {et~al.}(2018)\citenamefont {Ozawa},
  \citenamefont {Berthier}, \citenamefont {Biroli}, \citenamefont {Rosso},\
  and\ \citenamefont {Tarjus}}]{Tarjus_PNAS18}%
  \BibitemOpen
  \bibfield  {author} {\bibinfo {author} {\bibfnamefont {M.}~\bibnamefont
  {Ozawa}}, \bibinfo {author} {\bibfnamefont {L.}~\bibnamefont {Berthier}},
  \bibinfo {author} {\bibfnamefont {G.}~\bibnamefont {Biroli}}, \bibinfo
  {author} {\bibfnamefont {A.}~\bibnamefont {Rosso}}, \ and\ \bibinfo {author}
  {\bibfnamefont {G.}~\bibnamefont {Tarjus}},\ }\href {\doibase
  10.1073/pnas.1806156115} {\bibfield  {journal} {\bibinfo  {journal} {Proc.
  Natl. Acad. Sci. U.S.A.}\ }\textbf {\bibinfo {volume} {115}},\ \bibinfo
  {pages} {6656} (\bibinfo {year} {2018})}\BibitemShut {NoStop}%
\bibitem [{\citenamefont {Nandi}\ \emph {et~al.}(2016)\citenamefont {Nandi},
  \citenamefont {Biroli},\ and\ \citenamefont {Tarjus}}]{Nandi_prl16}%
  \BibitemOpen
  \bibfield  {author} {\bibinfo {author} {\bibfnamefont {S.~K.}\ \bibnamefont
  {Nandi}}, \bibinfo {author} {\bibfnamefont {G.}~\bibnamefont {Biroli}}, \
  and\ \bibinfo {author} {\bibfnamefont {G.}~\bibnamefont {Tarjus}},\ }\href
  {\doibase 10.1103/PhysRevLett.116.145701} {\bibfield  {journal} {\bibinfo
  {journal} {Phys. Rev. Lett.}\ }\textbf {\bibinfo {volume} {116}},\ \bibinfo
  {pages} {145701} (\bibinfo {year} {2016})}\BibitemShut {NoStop}%
\bibitem [{\citenamefont {Scheffer}\ \emph {et~al.}(2009)\citenamefont
  {Scheffer}, \citenamefont {Bascompte}, \citenamefont {Brock}, \citenamefont
  {Brovkin}, \citenamefont {Carpenter}, \citenamefont {Dakos}, \citenamefont
  {Held}, \citenamefont {Van~Nes}, \citenamefont {Rietkerk},\ and\
  \citenamefont {Sugihara}}]{Scheffer_nat09}%
  \BibitemOpen
  \bibfield  {author} {\bibinfo {author} {\bibfnamefont {M.}~\bibnamefont
  {Scheffer}}, \bibinfo {author} {\bibfnamefont {J.}~\bibnamefont {Bascompte}},
  \bibinfo {author} {\bibfnamefont {W.~A.}\ \bibnamefont {Brock}}, \bibinfo
  {author} {\bibfnamefont {V.}~\bibnamefont {Brovkin}}, \bibinfo {author}
  {\bibfnamefont {S.~R.}\ \bibnamefont {Carpenter}}, \bibinfo {author}
  {\bibfnamefont {V.}~\bibnamefont {Dakos}}, \bibinfo {author} {\bibfnamefont
  {H.}~\bibnamefont {Held}}, \bibinfo {author} {\bibfnamefont {E.~H.}\
  \bibnamefont {Van~Nes}}, \bibinfo {author} {\bibfnamefont {M.}~\bibnamefont
  {Rietkerk}}, \ and\ \bibinfo {author} {\bibfnamefont {G.}~\bibnamefont
  {Sugihara}},\ }\href {\doibase 10.1038/nature08227} {\bibfield  {journal}
  {\bibinfo  {journal} {Nature}\ }\textbf {\bibinfo {volume} {461}},\ \bibinfo
  {pages} {53} (\bibinfo {year} {2009})}\BibitemShut {NoStop}%
\bibitem [{\citenamefont {Scheffer}\ \emph {et~al.}(2012)\citenamefont
  {Scheffer}, \citenamefont {Carpenter}, \citenamefont {Lenton}, \citenamefont
  {Bascompte}, \citenamefont {Brock}, \citenamefont {Dakos}, \citenamefont
  {van~de Koppel}, \citenamefont {van~de Leemput}, \citenamefont {Levin},
  \citenamefont {van Nes}, \citenamefont {Pascual},\ and\ \citenamefont
  {Vandermeer}}]{Scheffer_sci12}%
  \BibitemOpen
  \bibfield  {author} {\bibinfo {author} {\bibfnamefont {M.}~\bibnamefont
  {Scheffer}}, \bibinfo {author} {\bibfnamefont {S.~R.}\ \bibnamefont
  {Carpenter}}, \bibinfo {author} {\bibfnamefont {T.~M.}\ \bibnamefont
  {Lenton}}, \bibinfo {author} {\bibfnamefont {J.}~\bibnamefont {Bascompte}},
  \bibinfo {author} {\bibfnamefont {W.}~\bibnamefont {Brock}}, \bibinfo
  {author} {\bibfnamefont {V.}~\bibnamefont {Dakos}}, \bibinfo {author}
  {\bibfnamefont {J.}~\bibnamefont {van~de Koppel}}, \bibinfo {author}
  {\bibfnamefont {I.~A.}\ \bibnamefont {van~de Leemput}}, \bibinfo {author}
  {\bibfnamefont {S.~A.}\ \bibnamefont {Levin}}, \bibinfo {author}
  {\bibfnamefont {E.~H.}\ \bibnamefont {van Nes}}, \bibinfo {author}
  {\bibfnamefont {M.}~\bibnamefont {Pascual}}, \ and\ \bibinfo {author}
  {\bibfnamefont {J.}~\bibnamefont {Vandermeer}},\ }\href {\doibase
  10.1126/science.1225244} {\bibfield  {journal} {\bibinfo  {journal}
  {Science}\ }\textbf {\bibinfo {volume} {338}},\ \bibinfo {pages} {344}
  (\bibinfo {year} {2012})}\BibitemShut {NoStop}%
\bibitem [{\citenamefont {Binder}(1984)}]{Binder_pra84}%
  \BibitemOpen
  \bibfield  {author} {\bibinfo {author} {\bibfnamefont {K.}~\bibnamefont
  {Binder}},\ }\href {\doibase 10.1103/PhysRevA.29.341} {\bibfield  {journal}
  {\bibinfo  {journal} {Phys. Rev. A}\ }\textbf {\bibinfo {volume} {29}},\
  \bibinfo {pages} {341} (\bibinfo {year} {1984})}\BibitemShut {NoStop}%
\bibitem [{\citenamefont {Zapperi}\ \emph {et~al.}(1997)\citenamefont
  {Zapperi}, \citenamefont {Ray}, \citenamefont {Stanley},\ and\ \citenamefont
  {Vespignani}}]{Zapperi_prl97}%
  \BibitemOpen
  \bibfield  {author} {\bibinfo {author} {\bibfnamefont {S.}~\bibnamefont
  {Zapperi}}, \bibinfo {author} {\bibfnamefont {P.}~\bibnamefont {Ray}},
  \bibinfo {author} {\bibfnamefont {H.~E.}\ \bibnamefont {Stanley}}, \ and\
  \bibinfo {author} {\bibfnamefont {A.}~\bibnamefont {Vespignani}},\ }\href
  {\doibase 10.1103/PhysRevLett.78.1408} {\bibfield  {journal} {\bibinfo
  {journal} {Phys. Rev. Lett.}\ }\textbf {\bibinfo {volume} {78}},\ \bibinfo
  {pages} {1408} (\bibinfo {year} {1997})}\BibitemShut {NoStop}%
\bibitem [{\citenamefont {Debenedetti}(2020)}]{PTBook_Metastable}%
  \BibitemOpen
  \bibfield  {author} {\bibinfo {author} {\bibfnamefont {P.}~\bibnamefont
  {Debenedetti}},\ }\href {https://books.google.co.in/books?id=WGvdDwAAQBAJ}
  {\emph {\bibinfo {title} {Metastable Liquids: Concepts and Principles}}},\
  Physical Chemistry: Science and Engineering\ (\bibinfo  {publisher}
  {Princeton University Press},\ \bibinfo {year} {2020})\BibitemShut {NoStop}%
\bibitem [{\citenamefont {P\'erez-Reche}\ \emph {et~al.}(2001)\citenamefont
  {P\'erez-Reche}, \citenamefont {Vives}, \citenamefont {Ma\~nosa},\ and\
  \citenamefont {Planes}}]{Planes_prl01}%
  \BibitemOpen
  \bibfield  {author} {\bibinfo {author} {\bibfnamefont {F.~J.}\ \bibnamefont
  {P\'erez-Reche}}, \bibinfo {author} {\bibfnamefont {E.}~\bibnamefont
  {Vives}}, \bibinfo {author} {\bibfnamefont {L.}~\bibnamefont {Ma\~nosa}}, \
  and\ \bibinfo {author} {\bibfnamefont {A.}~\bibnamefont {Planes}},\ }\href
  {\doibase 10.1103/PhysRevLett.87.195701} {\bibfield  {journal} {\bibinfo
  {journal} {Phys. Rev. Lett.}\ }\textbf {\bibinfo {volume} {87}},\ \bibinfo
  {pages} {195701} (\bibinfo {year} {2001})}\BibitemShut {NoStop}%
\bibitem [{\citenamefont {P\'erez-Reche}\ \emph
  {et~al.}(2004{\natexlab{a}})\citenamefont {P\'erez-Reche}, \citenamefont
  {Stipcich}, \citenamefont {Vives}, \citenamefont {Ma\~nosa}, \citenamefont
  {Planes},\ and\ \citenamefont {Morin}}]{Francisco_prb04}%
  \BibitemOpen
  \bibfield  {author} {\bibinfo {author} {\bibfnamefont {F.-J.}\ \bibnamefont
  {P\'erez-Reche}}, \bibinfo {author} {\bibfnamefont {M.}~\bibnamefont
  {Stipcich}}, \bibinfo {author} {\bibfnamefont {E.}~\bibnamefont {Vives}},
  \bibinfo {author} {\bibfnamefont {L.}~\bibnamefont {Ma\~nosa}}, \bibinfo
  {author} {\bibfnamefont {A.}~\bibnamefont {Planes}}, \ and\ \bibinfo {author}
  {\bibfnamefont {M.}~\bibnamefont {Morin}},\ }\href {\doibase
  10.1103/PhysRevB.69.064101} {\bibfield  {journal} {\bibinfo  {journal} {Phys.
  Rev. B}\ }\textbf {\bibinfo {volume} {69}},\ \bibinfo {pages} {064101}
  (\bibinfo {year} {2004}{\natexlab{a}})}\BibitemShut {NoStop}%
\bibitem [{\citenamefont {P\'erez-Reche}\ \emph {et~al.}(2007)\citenamefont
  {P\'erez-Reche}, \citenamefont {Truskinovsky},\ and\ \citenamefont
  {Zanzotto}}]{Francisco_prl07}%
  \BibitemOpen
  \bibfield  {author} {\bibinfo {author} {\bibfnamefont {F.-J.}\ \bibnamefont
  {P\'erez-Reche}}, \bibinfo {author} {\bibfnamefont {L.}~\bibnamefont
  {Truskinovsky}}, \ and\ \bibinfo {author} {\bibfnamefont {G.}~\bibnamefont
  {Zanzotto}},\ }\href {\doibase 10.1103/PhysRevLett.99.075501} {\bibfield
  {journal} {\bibinfo  {journal} {Phys. Rev. Lett.}\ }\textbf {\bibinfo
  {volume} {99}},\ \bibinfo {pages} {075501} (\bibinfo {year}
  {2007})}\BibitemShut {NoStop}%
\bibitem [{\citenamefont {Perez-Reche}\ \emph {et~al.}(2016)\citenamefont
  {Perez-Reche}, \citenamefont {Triguero}, \citenamefont {Zanzotto},\ and\
  \citenamefont {Truskinovsky}}]{Francisco_prb16}%
  \BibitemOpen
  \bibfield  {author} {\bibinfo {author} {\bibfnamefont {F.~J.}\ \bibnamefont
  {Perez-Reche}}, \bibinfo {author} {\bibfnamefont {C.}~\bibnamefont
  {Triguero}}, \bibinfo {author} {\bibfnamefont {G.}~\bibnamefont {Zanzotto}},
  \ and\ \bibinfo {author} {\bibfnamefont {L.}~\bibnamefont {Truskinovsky}},\
  }\href {\doibase 10.1103/PhysRevB.94.144102} {\bibfield  {journal} {\bibinfo
  {journal} {Phys. Rev. B}\ }\textbf {\bibinfo {volume} {94}},\ \bibinfo
  {pages} {144102} (\bibinfo {year} {2016})}\BibitemShut {NoStop}%
\bibitem [{\citenamefont {Liu}\ \emph {et~al.}(2016{\natexlab{a}})\citenamefont
  {Liu}, \citenamefont {Ferrero}, \citenamefont {Puosi}, \citenamefont
  {Barrat},\ and\ \citenamefont {Martens}}]{Liu_prl16}%
  \BibitemOpen
  \bibfield  {author} {\bibinfo {author} {\bibfnamefont {C.}~\bibnamefont
  {Liu}}, \bibinfo {author} {\bibfnamefont {E.~E.}\ \bibnamefont {Ferrero}},
  \bibinfo {author} {\bibfnamefont {F.}~\bibnamefont {Puosi}}, \bibinfo
  {author} {\bibfnamefont {J.-L.}\ \bibnamefont {Barrat}}, \ and\ \bibinfo
  {author} {\bibfnamefont {K.}~\bibnamefont {Martens}},\ }\href {\doibase
  10.1103/PhysRevLett.116.065501} {\bibfield  {journal} {\bibinfo  {journal}
  {Phys. Rev. Lett.}\ }\textbf {\bibinfo {volume} {116}},\ \bibinfo {pages}
  {065501} (\bibinfo {year} {2016}{\natexlab{a}})}\BibitemShut {NoStop}%
\bibitem [{\citenamefont {Cao}\ \emph {et~al.}(1990)\citenamefont {Cao},
  \citenamefont {Krumhansl},\ and\ \citenamefont {Gooding}}]{Cao_prb90}%
  \BibitemOpen
  \bibfield  {author} {\bibinfo {author} {\bibfnamefont {W.}~\bibnamefont
  {Cao}}, \bibinfo {author} {\bibfnamefont {J.~A.}\ \bibnamefont {Krumhansl}},
  \ and\ \bibinfo {author} {\bibfnamefont {R.~J.}\ \bibnamefont {Gooding}},\
  }\href {\doibase 10.1103/PhysRevB.41.11319} {\bibfield  {journal} {\bibinfo
  {journal} {Phys. Rev. B}\ }\textbf {\bibinfo {volume} {41}},\ \bibinfo
  {pages} {11319} (\bibinfo {year} {1990})}\BibitemShut {NoStop}%
\bibitem [{\citenamefont {Imry}\ and\ \citenamefont
  {Wortis}(1979)}]{Imry_prb79}%
  \BibitemOpen
  \bibfield  {author} {\bibinfo {author} {\bibfnamefont {Y.}~\bibnamefont
  {Imry}}\ and\ \bibinfo {author} {\bibfnamefont {M.}~\bibnamefont {Wortis}},\
  }\href {\doibase 10.1103/PhysRevB.19.3580} {\bibfield  {journal} {\bibinfo
  {journal} {Phys. Rev. B}\ }\textbf {\bibinfo {volume} {19}},\ \bibinfo
  {pages} {3580} (\bibinfo {year} {1979})}\BibitemShut {NoStop}%
\bibitem [{\citenamefont {Fan}\ \emph {et~al.}(2011)\citenamefont {Fan},
  \citenamefont {Cao}, \citenamefont {Seidel}, \citenamefont {Gu},
  \citenamefont {Yim}, \citenamefont {Barrett}, \citenamefont {Yu},
  \citenamefont {Ji}, \citenamefont {Ramesh}, \citenamefont {Chen},\ and\
  \citenamefont {Wu}}]{Fan_prb11}%
  \BibitemOpen
  \bibfield  {author} {\bibinfo {author} {\bibfnamefont {W.}~\bibnamefont
  {Fan}}, \bibinfo {author} {\bibfnamefont {J.}~\bibnamefont {Cao}}, \bibinfo
  {author} {\bibfnamefont {J.}~\bibnamefont {Seidel}}, \bibinfo {author}
  {\bibfnamefont {Y.}~\bibnamefont {Gu}}, \bibinfo {author} {\bibfnamefont
  {J.~W.}\ \bibnamefont {Yim}}, \bibinfo {author} {\bibfnamefont
  {C.}~\bibnamefont {Barrett}}, \bibinfo {author} {\bibfnamefont {K.~M.}\
  \bibnamefont {Yu}}, \bibinfo {author} {\bibfnamefont {J.}~\bibnamefont {Ji}},
  \bibinfo {author} {\bibfnamefont {R.}~\bibnamefont {Ramesh}}, \bibinfo
  {author} {\bibfnamefont {L.~Q.}\ \bibnamefont {Chen}}, \ and\ \bibinfo
  {author} {\bibfnamefont {J.}~\bibnamefont {Wu}},\ }\href {\doibase
  10.1103/PhysRevB.83.235102} {\bibfield  {journal} {\bibinfo  {journal} {Phys.
  Rev. B}\ }\textbf {\bibinfo {volume} {83}},\ \bibinfo {pages} {235102}
  (\bibinfo {year} {2011})}\BibitemShut {NoStop}%
\bibitem [{\citenamefont {Scheifele}\ \emph {et~al.}(2013)\citenamefont
  {Scheifele}, \citenamefont {Saika-Voivod}, \citenamefont {Bowles},\ and\
  \citenamefont {Poole}}]{Scheifele_pre13}%
  \BibitemOpen
  \bibfield  {author} {\bibinfo {author} {\bibfnamefont {B.}~\bibnamefont
  {Scheifele}}, \bibinfo {author} {\bibfnamefont {I.}~\bibnamefont
  {Saika-Voivod}}, \bibinfo {author} {\bibfnamefont {R.~K.}\ \bibnamefont
  {Bowles}}, \ and\ \bibinfo {author} {\bibfnamefont {P.~H.}\ \bibnamefont
  {Poole}},\ }\href {\doibase 10.1103/PhysRevE.87.042407} {\bibfield  {journal}
  {\bibinfo  {journal} {Phys. Rev. E}\ }\textbf {\bibinfo {volume} {87}},\
  \bibinfo {pages} {042407} (\bibinfo {year} {2013})}\BibitemShut {NoStop}%
\bibitem [{\citenamefont {Wang}\ \emph {et~al.}(2007)\citenamefont {Wang},
  \citenamefont {Gould},\ and\ \citenamefont {Klein}}]{Wang_pre07}%
  \BibitemOpen
  \bibfield  {author} {\bibinfo {author} {\bibfnamefont {H.}~\bibnamefont
  {Wang}}, \bibinfo {author} {\bibfnamefont {H.}~\bibnamefont {Gould}}, \ and\
  \bibinfo {author} {\bibfnamefont {W.}~\bibnamefont {Klein}},\ }\href
  {\doibase 10.1103/PhysRevE.76.031604} {\bibfield  {journal} {\bibinfo
  {journal} {Phys. Rev. E}\ }\textbf {\bibinfo {volume} {76}},\ \bibinfo
  {pages} {031604} (\bibinfo {year} {2007})}\BibitemShut {NoStop}%
\bibitem [{\citenamefont {Bhowmik}\ \emph {et~al.}(2019)\citenamefont
  {Bhowmik}, \citenamefont {Karmakar}, \citenamefont {Procaccia},\ and\
  \citenamefont {Rainone}}]{Bhowmik_pre19}%
  \BibitemOpen
  \bibfield  {author} {\bibinfo {author} {\bibfnamefont {B.~P.}\ \bibnamefont
  {Bhowmik}}, \bibinfo {author} {\bibfnamefont {S.}~\bibnamefont {Karmakar}},
  \bibinfo {author} {\bibfnamefont {I.}~\bibnamefont {Procaccia}}, \ and\
  \bibinfo {author} {\bibfnamefont {C.}~\bibnamefont {Rainone}},\ }\href
  {\doibase 10.1103/PhysRevE.100.052110} {\bibfield  {journal} {\bibinfo
  {journal} {Phys. Rev. E}\ }\textbf {\bibinfo {volume} {100}},\ \bibinfo
  {pages} {052110} (\bibinfo {year} {2019})}\BibitemShut {NoStop}%
\bibitem [{\citenamefont {Sethna}\ \emph {et~al.}(1993)\citenamefont {Sethna},
  \citenamefont {Dahmen}, \citenamefont {Kartha}, \citenamefont {Krumhansl},
  \citenamefont {Roberts},\ and\ \citenamefont {Shore}}]{Sethna_prl93}%
  \BibitemOpen
  \bibfield  {author} {\bibinfo {author} {\bibfnamefont {J.~P.}\ \bibnamefont
  {Sethna}}, \bibinfo {author} {\bibfnamefont {K.}~\bibnamefont {Dahmen}},
  \bibinfo {author} {\bibfnamefont {S.}~\bibnamefont {Kartha}}, \bibinfo
  {author} {\bibfnamefont {J.~A.}\ \bibnamefont {Krumhansl}}, \bibinfo {author}
  {\bibfnamefont {B.~W.}\ \bibnamefont {Roberts}}, \ and\ \bibinfo {author}
  {\bibfnamefont {J.~D.}\ \bibnamefont {Shore}},\ }\href {\doibase
  10.1103/PhysRevLett.70.3347} {\bibfield  {journal} {\bibinfo  {journal}
  {Phys. Rev. Lett.}\ }\textbf {\bibinfo {volume} {70}},\ \bibinfo {pages}
  {3347} (\bibinfo {year} {1993})}\BibitemShut {NoStop}%
\bibitem [{\citenamefont {Perkovi\ifmmode~\acute{c}\else \'{c}\fi{}}\ \emph
  {et~al.}(1995)\citenamefont {Perkovi\ifmmode~\acute{c}\else \'{c}\fi{}},
  \citenamefont {Dahmen},\ and\ \citenamefont {Sethna}}]{Sethna_prl95}%
  \BibitemOpen
  \bibfield  {author} {\bibinfo {author} {\bibfnamefont {O.}~\bibnamefont
  {Perkovi\ifmmode~\acute{c}\else \'{c}\fi{}}}, \bibinfo {author}
  {\bibfnamefont {K.}~\bibnamefont {Dahmen}}, \ and\ \bibinfo {author}
  {\bibfnamefont {J.~P.}\ \bibnamefont {Sethna}},\ }\href {\doibase
  10.1103/PhysRevLett.75.4528} {\bibfield  {journal} {\bibinfo  {journal}
  {Phys. Rev. Lett.}\ }\textbf {\bibinfo {volume} {75}},\ \bibinfo {pages}
  {4528} (\bibinfo {year} {1995})}\BibitemShut {NoStop}%
\bibitem [{\citenamefont {Pierce}\ \emph {et~al.}(2007)\citenamefont {Pierce},
  \citenamefont {Buechler}, \citenamefont {Sorensen}, \citenamefont {Kevan},
  \citenamefont {Jagla}, \citenamefont {Deutsch}, \citenamefont {Mai},
  \citenamefont {Narayan}, \citenamefont {Davies}, \citenamefont {Liu},
  \citenamefont {Zimanyi}, \citenamefont {Katzgraber}, \citenamefont {Hellwig},
  \citenamefont {Fullerton}, \citenamefont {Fischer},\ and\ \citenamefont
  {Kortright}}]{Pierce_prb07}%
  \BibitemOpen
  \bibfield  {author} {\bibinfo {author} {\bibfnamefont {M.~S.}\ \bibnamefont
  {Pierce}}, \bibinfo {author} {\bibfnamefont {C.~R.}\ \bibnamefont
  {Buechler}}, \bibinfo {author} {\bibfnamefont {L.~B.}\ \bibnamefont
  {Sorensen}}, \bibinfo {author} {\bibfnamefont {S.~D.}\ \bibnamefont {Kevan}},
  \bibinfo {author} {\bibfnamefont {E.~A.}\ \bibnamefont {Jagla}}, \bibinfo
  {author} {\bibfnamefont {J.~M.}\ \bibnamefont {Deutsch}}, \bibinfo {author}
  {\bibfnamefont {T.}~\bibnamefont {Mai}}, \bibinfo {author} {\bibfnamefont
  {O.}~\bibnamefont {Narayan}}, \bibinfo {author} {\bibfnamefont {J.~E.}\
  \bibnamefont {Davies}}, \bibinfo {author} {\bibfnamefont {K.}~\bibnamefont
  {Liu}}, \bibinfo {author} {\bibfnamefont {G.~T.}\ \bibnamefont {Zimanyi}},
  \bibinfo {author} {\bibfnamefont {H.~G.}\ \bibnamefont {Katzgraber}},
  \bibinfo {author} {\bibfnamefont {O.}~\bibnamefont {Hellwig}}, \bibinfo
  {author} {\bibfnamefont {E.~E.}\ \bibnamefont {Fullerton}}, \bibinfo {author}
  {\bibfnamefont {P.}~\bibnamefont {Fischer}}, \ and\ \bibinfo {author}
  {\bibfnamefont {J.~B.}\ \bibnamefont {Kortright}},\ }\href {\doibase
  10.1103/PhysRevB.75.144406} {\bibfield  {journal} {\bibinfo  {journal} {Phys.
  Rev. B}\ }\textbf {\bibinfo {volume} {75}},\ \bibinfo {pages} {144406}
  (\bibinfo {year} {2007})}\BibitemShut {NoStop}%
\bibitem [{\citenamefont {Lee}\ \emph {et~al.}(2016)\citenamefont {Lee},
  \citenamefont {Kim}, \citenamefont {Hwang}, \citenamefont {Noh},\ and\
  \citenamefont {Jhe}}]{Lee_pre16}%
  \BibitemOpen
  \bibfield  {author} {\bibinfo {author} {\bibfnamefont {W.}~\bibnamefont
  {Lee}}, \bibinfo {author} {\bibfnamefont {J.-H.}\ \bibnamefont {Kim}},
  \bibinfo {author} {\bibfnamefont {J.~G.}\ \bibnamefont {Hwang}}, \bibinfo
  {author} {\bibfnamefont {H.-R.}\ \bibnamefont {Noh}}, \ and\ \bibinfo
  {author} {\bibfnamefont {W.}~\bibnamefont {Jhe}},\ }\href {\doibase
  10.1103/PhysRevE.94.032141} {\bibfield  {journal} {\bibinfo  {journal} {Phys.
  Rev. E}\ }\textbf {\bibinfo {volume} {94}},\ \bibinfo {pages} {032141}
  (\bibinfo {year} {2016})}\BibitemShut {NoStop}%
\bibitem [{\citenamefont {Y\ifmmode \imath \else \i \fi{}ld\ifmmode \imath
  \else~\i \fi{}z}\ \emph {et~al.}(2004)\citenamefont {Y\ifmmode \imath \else
  \i \fi{}ld\ifmmode \imath \else~\i \fi{}z}, \citenamefont {Pekcan},
  \citenamefont {Berker},\ and\ \citenamefont {\"Ozbek}}]{Yildiz_pre04}%
  \BibitemOpen
  \bibfield  {author} {\bibinfo {author} {\bibfnamefont {S.}~\bibnamefont
  {Y\ifmmode \imath \else \i \fi{}ld\ifmmode \imath \else~\i \fi{}z}}, \bibinfo
  {author} {\bibfnamefont {O.}~\bibnamefont {Pekcan}}, \bibinfo {author}
  {\bibfnamefont {A.~N.}\ \bibnamefont {Berker}}, \ and\ \bibinfo {author}
  {\bibfnamefont {H.}~\bibnamefont {\"Ozbek}},\ }\href {\doibase
  10.1103/PhysRevE.69.031705} {\bibfield  {journal} {\bibinfo  {journal} {Phys.
  Rev. E}\ }\textbf {\bibinfo {volume} {69}},\ \bibinfo {pages} {031705}
  (\bibinfo {year} {2004})}\BibitemShut {NoStop}%
\bibitem [{\citenamefont {Wang}\ \emph {et~al.}(2011)\citenamefont {Wang},
  \citenamefont {Li},\ and\ \citenamefont {Zhang}}]{Wang_JCP11}%
  \BibitemOpen
  \bibfield  {author} {\bibinfo {author} {\bibfnamefont {Y.-Z.}\ \bibnamefont
  {Wang}}, \bibinfo {author} {\bibfnamefont {Y.}~\bibnamefont {Li}}, \ and\
  \bibinfo {author} {\bibfnamefont {J.-X.}\ \bibnamefont {Zhang}},\ }\href
  {\doibase 10.1063/1.3564919} {\bibfield  {journal} {\bibinfo  {journal} {J.
  Chem. Phys.}\ }\textbf {\bibinfo {volume} {134}},\ \bibinfo {pages} {114510}
  (\bibinfo {year} {2011})}\BibitemShut {NoStop}%
\bibitem [{\citenamefont {He}\ and\ \citenamefont {Wang}(1993)}]{He_prl93}%
  \BibitemOpen
  \bibfield  {author} {\bibinfo {author} {\bibfnamefont {Y.-L.}\ \bibnamefont
  {He}}\ and\ \bibinfo {author} {\bibfnamefont {G.-C.}\ \bibnamefont {Wang}},\
  }\href {\doibase 10.1103/PhysRevLett.70.2336} {\bibfield  {journal} {\bibinfo
   {journal} {Phys. Rev. Lett.}\ }\textbf {\bibinfo {volume} {70}},\ \bibinfo
  {pages} {2336} (\bibinfo {year} {1993})}\BibitemShut {NoStop}%
\bibitem [{\citenamefont {Jiang}\ \emph {et~al.}(1995)\citenamefont {Jiang},
  \citenamefont {Yang},\ and\ \citenamefont {Wang}}]{Jiang_prb95}%
  \BibitemOpen
  \bibfield  {author} {\bibinfo {author} {\bibfnamefont {Q.}~\bibnamefont
  {Jiang}}, \bibinfo {author} {\bibfnamefont {H.-N.}\ \bibnamefont {Yang}}, \
  and\ \bibinfo {author} {\bibfnamefont {G.-C.}\ \bibnamefont {Wang}},\ }\href
  {\doibase 10.1103/PhysRevB.52.14911} {\bibfield  {journal} {\bibinfo
  {journal} {Phys. Rev. B}\ }\textbf {\bibinfo {volume} {52}},\ \bibinfo
  {pages} {14911} (\bibinfo {year} {1995})}\BibitemShut {NoStop}%
\bibitem [{\citenamefont {Pan}\ \emph {et~al.}(2003{\natexlab{a}})\citenamefont
  {Pan}, \citenamefont {Yu}, \citenamefont {Wu}, \citenamefont {Zhou},\ and\
  \citenamefont {Liu}}]{Pan_apl03}%
  \BibitemOpen
  \bibfield  {author} {\bibinfo {author} {\bibfnamefont {B.}~\bibnamefont
  {Pan}}, \bibinfo {author} {\bibfnamefont {H.}~\bibnamefont {Yu}}, \bibinfo
  {author} {\bibfnamefont {D.}~\bibnamefont {Wu}}, \bibinfo {author}
  {\bibfnamefont {X.~H.}\ \bibnamefont {Zhou}}, \ and\ \bibinfo {author}
  {\bibfnamefont {J.~M.}\ \bibnamefont {Liu}},\ }\href {\doibase
  10.1063/1.1602580} {\bibfield  {journal} {\bibinfo  {journal} {Appl. Phys.
  Lett.}\ }\textbf {\bibinfo {volume} {83}},\ \bibinfo {pages} {1406} (\bibinfo
  {year} {2003}{\natexlab{a}})}\BibitemShut {NoStop}%
\bibitem [{\citenamefont {Liu}\ \emph {et~al.}(1999)\citenamefont {Liu},
  \citenamefont {Li}, \citenamefont {Ong},\ and\ \citenamefont
  {Lim}}]{Liu_jap99}%
  \BibitemOpen
  \bibfield  {author} {\bibinfo {author} {\bibfnamefont {J.-M.}\ \bibnamefont
  {Liu}}, \bibinfo {author} {\bibfnamefont {H.~P.}\ \bibnamefont {Li}},
  \bibinfo {author} {\bibfnamefont {C.~K.}\ \bibnamefont {Ong}}, \ and\
  \bibinfo {author} {\bibfnamefont {L.~C.}\ \bibnamefont {Lim}},\ }\href
  {\doibase 10.1063/1.371500} {\bibfield  {journal} {\bibinfo  {journal} {J.
  Appl. Phys.}\ }\textbf {\bibinfo {volume} {86}},\ \bibinfo {pages} {5198}
  (\bibinfo {year} {1999})}\BibitemShut {NoStop}%
\bibitem [{\citenamefont {Kim}\ and\ \citenamefont {Kim}(1997)}]{Kim_prb97}%
  \BibitemOpen
  \bibfield  {author} {\bibinfo {author} {\bibfnamefont {Y.-H.}\ \bibnamefont
  {Kim}}\ and\ \bibinfo {author} {\bibfnamefont {J.-J.}\ \bibnamefont {Kim}},\
  }\href {\doibase 10.1103/PhysRevB.55.R11933} {\bibfield  {journal} {\bibinfo
  {journal} {Phys. Rev. B}\ }\textbf {\bibinfo {volume} {55}},\ \bibinfo
  {pages} {R11933} (\bibinfo {year} {1997})}\BibitemShut {NoStop}%
\bibitem [{\citenamefont {Jung}\ \emph {et~al.}(1990)\citenamefont {Jung},
  \citenamefont {Gray}, \citenamefont {Roy},\ and\ \citenamefont
  {Mandel}}]{Jung_prl90}%
  \BibitemOpen
  \bibfield  {author} {\bibinfo {author} {\bibfnamefont {P.}~\bibnamefont
  {Jung}}, \bibinfo {author} {\bibfnamefont {G.}~\bibnamefont {Gray}}, \bibinfo
  {author} {\bibfnamefont {R.}~\bibnamefont {Roy}}, \ and\ \bibinfo {author}
  {\bibfnamefont {P.}~\bibnamefont {Mandel}},\ }\href {\doibase
  10.1103/PhysRevLett.65.1873} {\bibfield  {journal} {\bibinfo  {journal}
  {Phys. Rev. Lett.}\ }\textbf {\bibinfo {volume} {65}},\ \bibinfo {pages}
  {1873} (\bibinfo {year} {1990})}\BibitemShut {NoStop}%
\bibitem [{\citenamefont {Kuang}\ \emph {et~al.}(2000)\citenamefont {Kuang},
  \citenamefont {Zhang}, \citenamefont {Zhang}, \citenamefont {Liang},\ and\
  \citenamefont {Fung}}]{Kuang_ssc00}%
  \BibitemOpen
  \bibfield  {author} {\bibinfo {author} {\bibfnamefont {Z.}~\bibnamefont
  {Kuang}}, \bibinfo {author} {\bibfnamefont {J.}~\bibnamefont {Zhang}},
  \bibinfo {author} {\bibfnamefont {X.}~\bibnamefont {Zhang}}, \bibinfo
  {author} {\bibfnamefont {K.}~\bibnamefont {Liang}}, \ and\ \bibinfo {author}
  {\bibfnamefont {P.}~\bibnamefont {Fung}},\ }\href {\doibase
  https://doi.org/10.1016/S0038-1098(00)00028-4} {\bibfield  {journal}
  {\bibinfo  {journal} {Solid State Commun.}\ }\textbf {\bibinfo {volume}
  {114}},\ \bibinfo {pages} {231} (\bibinfo {year} {2000})}\BibitemShut
  {NoStop}%
\bibitem [{\citenamefont {Wongdamnern}\ \emph {et~al.}(2009)\citenamefont
  {Wongdamnern}, \citenamefont {Ngamjarurojana}, \citenamefont {Laosiritaworn},
  \citenamefont {Ananta},\ and\ \citenamefont {Yimnirun}}]{Wongdamnern_jap09}%
  \BibitemOpen
  \bibfield  {author} {\bibinfo {author} {\bibfnamefont {N.}~\bibnamefont
  {Wongdamnern}}, \bibinfo {author} {\bibfnamefont {A.}~\bibnamefont
  {Ngamjarurojana}}, \bibinfo {author} {\bibfnamefont {Y.}~\bibnamefont
  {Laosiritaworn}}, \bibinfo {author} {\bibfnamefont {S.}~\bibnamefont
  {Ananta}}, \ and\ \bibinfo {author} {\bibfnamefont {R.}~\bibnamefont
  {Yimnirun}},\ }\href {\doibase 10.1063/1.3086317} {\bibfield  {journal}
  {\bibinfo  {journal} {J. Appl. Phys.}\ }\textbf {\bibinfo {volume} {105}},\
  \bibinfo {pages} {044109} (\bibinfo {year} {2009})}\BibitemShut {NoStop}%
\bibitem [{\citenamefont {Wongdamnern}\ \emph
  {et~al.}(2010{\natexlab{a}})\citenamefont {Wongdamnern}, \citenamefont
  {Tangsritragul}, \citenamefont {Ngamjarurojana}, \citenamefont {Ananta},
  \citenamefont {Laosiritaworn},\ and\ \citenamefont
  {Yimnirun}}]{Wongdamnern_mcp10}%
  \BibitemOpen
  \bibfield  {author} {\bibinfo {author} {\bibfnamefont {N.}~\bibnamefont
  {Wongdamnern}}, \bibinfo {author} {\bibfnamefont {J.}~\bibnamefont
  {Tangsritragul}}, \bibinfo {author} {\bibfnamefont {A.}~\bibnamefont
  {Ngamjarurojana}}, \bibinfo {author} {\bibfnamefont {S.}~\bibnamefont
  {Ananta}}, \bibinfo {author} {\bibfnamefont {Y.}~\bibnamefont
  {Laosiritaworn}}, \ and\ \bibinfo {author} {\bibfnamefont {R.}~\bibnamefont
  {Yimnirun}},\ }\href {\doibase
  https://doi.org/10.1016/j.matchemphys.2010.06.032} {\bibfield  {journal}
  {\bibinfo  {journal} {Materials Chemistry and Physics}\ }\textbf {\bibinfo
  {volume} {124}},\ \bibinfo {pages} {281} (\bibinfo {year}
  {2010}{\natexlab{a}})}\BibitemShut {NoStop}%
\bibitem [{\citenamefont {Wongdamnern}\ \emph
  {et~al.}(2010{\natexlab{b}})\citenamefont {Wongdamnern}, \citenamefont
  {Ngamjarurojana}, \citenamefont {Ananta}, \citenamefont {Laosiritaworn},\
  and\ \citenamefont {Yimnirun}}]{wongdamnern_kem10}%
  \BibitemOpen
  \bibfield  {author} {\bibinfo {author} {\bibfnamefont {N.}~\bibnamefont
  {Wongdamnern}}, \bibinfo {author} {\bibfnamefont {A.}~\bibnamefont
  {Ngamjarurojana}}, \bibinfo {author} {\bibfnamefont {S.}~\bibnamefont
  {Ananta}}, \bibinfo {author} {\bibfnamefont {Y.}~\bibnamefont
  {Laosiritaworn}}, \ and\ \bibinfo {author} {\bibfnamefont {R.}~\bibnamefont
  {Yimnirun}},\ }in\ \href {\doibase
  10.4028/www.scientific.net/KEM.421-422.399} {\emph {\bibinfo {booktitle}
  {Asian Ceramic Science for Electronics III and Electroceramics in Japan
  XII}}},\ \bibinfo {series} {Key Engineering Materials}, Vol.\ \bibinfo
  {volume} {421}\ (\bibinfo  {publisher} {Trans Tech Publications Ltd},\
  \bibinfo {year} {2010})\ pp.\ \bibinfo {pages} {399--402}\BibitemShut
  {NoStop}%
\bibitem [{\citenamefont {Yimnirun}\ \emph {et~al.}(2006)\citenamefont
  {Yimnirun}, \citenamefont {Laosiritaworn}, \citenamefont {Wongsaenmai},\ and\
  \citenamefont {Ananta}}]{Yimnirun_apl06}%
  \BibitemOpen
  \bibfield  {author} {\bibinfo {author} {\bibfnamefont {R.}~\bibnamefont
  {Yimnirun}}, \bibinfo {author} {\bibfnamefont {Y.}~\bibnamefont
  {Laosiritaworn}}, \bibinfo {author} {\bibfnamefont {S.}~\bibnamefont
  {Wongsaenmai}}, \ and\ \bibinfo {author} {\bibfnamefont {S.}~\bibnamefont
  {Ananta}},\ }\href {\doibase 10.1063/1.2363143} {\bibfield  {journal}
  {\bibinfo  {journal} {Appl. Phys. Lett.}\ }\textbf {\bibinfo {volume} {89}},\
  \bibinfo {pages} {162901} (\bibinfo {year} {2006})}\BibitemShut {NoStop}%
\bibitem [{\citenamefont {Yimnirun}\ \emph {et~al.}(2007)\citenamefont
  {Yimnirun}, \citenamefont {Wongmaneerung}, \citenamefont {Wongsaenmai},
  \citenamefont {Ngamjarurojana}, \citenamefont {Ananta},\ and\ \citenamefont
  {Laosiritaworn}}]{Yimnirun_apl07}%
  \BibitemOpen
  \bibfield  {author} {\bibinfo {author} {\bibfnamefont {R.}~\bibnamefont
  {Yimnirun}}, \bibinfo {author} {\bibfnamefont {R.}~\bibnamefont
  {Wongmaneerung}}, \bibinfo {author} {\bibfnamefont {S.}~\bibnamefont
  {Wongsaenmai}}, \bibinfo {author} {\bibfnamefont {A.}~\bibnamefont
  {Ngamjarurojana}}, \bibinfo {author} {\bibfnamefont {S.}~\bibnamefont
  {Ananta}}, \ and\ \bibinfo {author} {\bibfnamefont {Y.}~\bibnamefont
  {Laosiritaworn}},\ }\href {\doibase 10.1063/1.2713769} {\bibfield  {journal}
  {\bibinfo  {journal} {Appl. Phys. Lett.}\ }\textbf {\bibinfo {volume} {90}},\
  \bibinfo {pages} {112908} (\bibinfo {year} {2007})}\BibitemShut {NoStop}%
\bibitem [{\citenamefont {Zhang}\ \emph {et~al.}(1996)\citenamefont {Zhang},
  \citenamefont {Zhong},\ and\ \citenamefont {Siu}}]{Zhang_ssc96}%
  \BibitemOpen
  \bibfield  {author} {\bibinfo {author} {\bibfnamefont {J.}~\bibnamefont
  {Zhang}}, \bibinfo {author} {\bibfnamefont {F.}~\bibnamefont {Zhong}}, \ and\
  \bibinfo {author} {\bibfnamefont {G.}~\bibnamefont {Siu}},\ }\href {\doibase
  https://doi.org/10.1016/0038-1098(95)00781-4} {\bibfield  {journal} {\bibinfo
   {journal} {Solid State Commun.}\ }\textbf {\bibinfo {volume} {97}},\
  \bibinfo {pages} {847} (\bibinfo {year} {1996})}\BibitemShut {NoStop}%
\bibitem [{\citenamefont {Pan}\ \emph {et~al.}(2003{\natexlab{b}})\citenamefont
  {Pan}, \citenamefont {Yang}, \citenamefont {Yu}, \citenamefont {Liu},
  \citenamefont {Li}, \citenamefont {Liu},\ and\ \citenamefont
  {Chan}}]{Pan_msc03}%
  \BibitemOpen
  \bibfield  {author} {\bibinfo {author} {\bibfnamefont {B.}~\bibnamefont
  {Pan}}, \bibinfo {author} {\bibfnamefont {Y.}~\bibnamefont {Yang}}, \bibinfo
  {author} {\bibfnamefont {L.-C.}\ \bibnamefont {Yu}}, \bibinfo {author}
  {\bibfnamefont {J.-M.}\ \bibnamefont {Liu}}, \bibinfo {author} {\bibfnamefont
  {K.}~\bibnamefont {Li}}, \bibinfo {author} {\bibfnamefont {Z.~G.}\
  \bibnamefont {Liu}}, \ and\ \bibinfo {author} {\bibfnamefont {H.~L.~W.}\
  \bibnamefont {Chan}},\ }\href {\doibase
  https://doi.org/10.1016/S0921-5107(02)00482-8} {\bibfield  {journal}
  {\bibinfo  {journal} {Materials Science and Engineering: B}\ }\textbf
  {\bibinfo {volume} {99}},\ \bibinfo {pages} {179} (\bibinfo {year}
  {2003}{\natexlab{b}})}\BibitemShut {NoStop}%
\bibitem [{\citenamefont {Prajapati}\ \emph {et~al.}(2022)\citenamefont
  {Prajapati}, \citenamefont {Kundu}, \citenamefont {Das}, \citenamefont {V},\
  and\ \citenamefont {Rana}}]{Prajapati_NJP22}%
  \BibitemOpen
  \bibfield  {author} {\bibinfo {author} {\bibfnamefont {G.~L.}\ \bibnamefont
  {Prajapati}}, \bibinfo {author} {\bibfnamefont {S.}~\bibnamefont {Kundu}},
  \bibinfo {author} {\bibfnamefont {S.}~\bibnamefont {Das}}, \bibinfo {author}
  {\bibfnamefont {T.~D.~V.}\ \bibnamefont {V}}, \ and\ \bibinfo {author}
  {\bibfnamefont {D.~S.}\ \bibnamefont {Rana}},\ }\href {\doibase
  10.1088/1367-2630/ac95b8} {\bibfield  {journal} {\bibinfo  {journal} {New
  Journal of Physics}\ }\textbf {\bibinfo {volume} {24}},\ \bibinfo {pages}
  {103016} (\bibinfo {year} {2022})}\BibitemShut {NoStop}%
\bibitem [{\citenamefont {Rao}\ \emph {et~al.}(1990)\citenamefont {Rao},
  \citenamefont {Krishnamurthy},\ and\ \citenamefont {Pandit}}]{Rao_prb90}%
  \BibitemOpen
  \bibfield  {author} {\bibinfo {author} {\bibfnamefont {M.}~\bibnamefont
  {Rao}}, \bibinfo {author} {\bibfnamefont {H.~R.}\ \bibnamefont
  {Krishnamurthy}}, \ and\ \bibinfo {author} {\bibfnamefont {R.}~\bibnamefont
  {Pandit}},\ }\href {\doibase 10.1103/PhysRevB.42.856} {\bibfield  {journal}
  {\bibinfo  {journal} {Phys. Rev. B}\ }\textbf {\bibinfo {volume} {42}},\
  \bibinfo {pages} {856} (\bibinfo {year} {1990})}\BibitemShut {NoStop}%
\bibitem [{\citenamefont {Rao}\ and\ \citenamefont {Pandit}(1991)}]{Rao_prb91}%
  \BibitemOpen
  \bibfield  {author} {\bibinfo {author} {\bibfnamefont {M.}~\bibnamefont
  {Rao}}\ and\ \bibinfo {author} {\bibfnamefont {R.}~\bibnamefont {Pandit}},\
  }\href {\doibase 10.1103/PhysRevB.43.3373} {\bibfield  {journal} {\bibinfo
  {journal} {Phys. Rev. B}\ }\textbf {\bibinfo {volume} {43}},\ \bibinfo
  {pages} {3373} (\bibinfo {year} {1991})}\BibitemShut {NoStop}%
\bibitem [{\citenamefont {Rao}(1992)}]{RaoComment_prl92}%
  \BibitemOpen
  \bibfield  {author} {\bibinfo {author} {\bibfnamefont {M.}~\bibnamefont
  {Rao}},\ }\href {\doibase 10.1103/PhysRevLett.68.1436} {\bibfield  {journal}
  {\bibinfo  {journal} {Phys. Rev. Lett.}\ }\textbf {\bibinfo {volume} {68}},\
  \bibinfo {pages} {1436} (\bibinfo {year} {1992})}\BibitemShut {NoStop}%
\bibitem [{\citenamefont {Liang}\ and\ \citenamefont
  {Zhong}(2017)}]{Zhong_FrontPhys17_Temp}%
  \BibitemOpen
  \bibfield  {author} {\bibinfo {author} {\bibfnamefont {N.}~\bibnamefont
  {Liang}}\ and\ \bibinfo {author} {\bibfnamefont {F.}~\bibnamefont {Zhong}},\
  }\href {\doibase 10.1007/s11467-016-0633-y} {\bibfield  {journal} {\bibinfo
  {journal} {Front. Phys.}\ }\textbf {\bibinfo {volume} {12}},\ \bibinfo
  {pages} {126403} (\bibinfo {year} {2017})}\BibitemShut {NoStop}%
\bibitem [{\citenamefont {Zhong}(2017)}]{Zhong_FrontPhys17_Field}%
  \BibitemOpen
  \bibfield  {author} {\bibinfo {author} {\bibfnamefont {F.}~\bibnamefont
  {Zhong}},\ }\href {\doibase 10.1007/s11467-016-0632-z} {\bibfield  {journal}
  {\bibinfo  {journal} {Front. Phys.}\ }\textbf {\bibinfo {volume} {12}},\
  \bibinfo {pages} {1} (\bibinfo {year} {2017})}\BibitemShut {NoStop}%
\bibitem [{\citenamefont {Zhong}\ and\ \citenamefont
  {Zhang}(1995)}]{Zhong_prl95}%
  \BibitemOpen
  \bibfield  {author} {\bibinfo {author} {\bibfnamefont {F.}~\bibnamefont
  {Zhong}}\ and\ \bibinfo {author} {\bibfnamefont {J.}~\bibnamefont {Zhang}},\
  }\href {\doibase 10.1103/PhysRevLett.75.2027} {\bibfield  {journal} {\bibinfo
   {journal} {Phys. Rev. Lett.}\ }\textbf {\bibinfo {volume} {75}},\ \bibinfo
  {pages} {2027} (\bibinfo {year} {1995})}\BibitemShut {NoStop}%
\bibitem [{\citenamefont {Zhong}\ and\ \citenamefont
  {Chen}(2005)}]{Zhong_prl05}%
  \BibitemOpen
  \bibfield  {author} {\bibinfo {author} {\bibfnamefont {F.}~\bibnamefont
  {Zhong}}\ and\ \bibinfo {author} {\bibfnamefont {Q.}~\bibnamefont {Chen}},\
  }\href {\doibase 10.1103/PhysRevLett.95.175701} {\bibfield  {journal}
  {\bibinfo  {journal} {Phys. Rev. Lett.}\ }\textbf {\bibinfo {volume} {95}},\
  \bibinfo {pages} {175701} (\bibinfo {year} {2005})}\BibitemShut {NoStop}%
\bibitem [{\citenamefont {Shukla}(2018)}]{Shukla_pre18}%
  \BibitemOpen
  \bibfield  {author} {\bibinfo {author} {\bibfnamefont {P.}~\bibnamefont
  {Shukla}},\ }\href {\doibase 10.1103/PhysRevE.97.062127} {\bibfield
  {journal} {\bibinfo  {journal} {Phys. Rev. E}\ }\textbf {\bibinfo {volume}
  {97}},\ \bibinfo {pages} {062127} (\bibinfo {year} {2018})}\BibitemShut
  {NoStop}%
\bibitem [{\citenamefont {Bray}(2002)}]{Bray_AP02}%
  \BibitemOpen
  \bibfield  {author} {\bibinfo {author} {\bibfnamefont {A.~J.}\ \bibnamefont
  {Bray}},\ }\href {\doibase 10.1080/00018730110117433} {\bibfield  {journal}
  {\bibinfo  {journal} {Advances in Physics}\ }\textbf {\bibinfo {volume}
  {51}},\ \bibinfo {pages} {481} (\bibinfo {year} {2002})}\BibitemShut
  {NoStop}%
\bibitem [{\citenamefont {Jani{\'{c}}evi{\'{c}}}\ \emph
  {et~al.}(2021)\citenamefont {Jani{\'{c}}evi{\'{c}}}, \citenamefont
  {Kne{\v{z}}evi{\'{c}}}, \citenamefont {Mijatovi{\'{c}}},\ and\ \citenamefont
  {Spasojevi{\'{c}}}}]{Sanja_JSM21}%
  \BibitemOpen
  \bibfield  {author} {\bibinfo {author} {\bibfnamefont {S.}~\bibnamefont
  {Jani{\'{c}}evi{\'{c}}}}, \bibinfo {author} {\bibfnamefont {D.}~\bibnamefont
  {Kne{\v{z}}evi{\'{c}}}}, \bibinfo {author} {\bibfnamefont {S.}~\bibnamefont
  {Mijatovi{\'{c}}}}, \ and\ \bibinfo {author} {\bibfnamefont {D.}~\bibnamefont
  {Spasojevi{\'{c}}}},\ }\href {\doibase 10.1088/1742-5468/abcd32} {\bibfield
  {journal} {\bibinfo  {journal} {Journal of Statistical Mechanics: Theory and
  Experiment}\ }\textbf {\bibinfo {volume} {2021}},\ \bibinfo {pages} {013202}
  (\bibinfo {year} {2021})}\BibitemShut {NoStop}%
\bibitem [{\citenamefont {Dahmen}\ and\ \citenamefont
  {Sethna}(1996)}]{Sethna_prb96}%
  \BibitemOpen
  \bibfield  {author} {\bibinfo {author} {\bibfnamefont {K.}~\bibnamefont
  {Dahmen}}\ and\ \bibinfo {author} {\bibfnamefont {J.~P.}\ \bibnamefont
  {Sethna}},\ }\href {\doibase 10.1103/PhysRevB.53.14872} {\bibfield  {journal}
  {\bibinfo  {journal} {Phys. Rev. B}\ }\textbf {\bibinfo {volume} {53}},\
  \bibinfo {pages} {14872} (\bibinfo {year} {1996})}\BibitemShut {NoStop}%
\bibitem [{\citenamefont {Chaikin}\ and\ \citenamefont
  {Lubensky}(2000)}]{CMBook_Lubensky}%
  \BibitemOpen
  \bibfield  {author} {\bibinfo {author} {\bibfnamefont {P.~M.}\ \bibnamefont
  {Chaikin}}\ and\ \bibinfo {author} {\bibfnamefont {T.~C.}\ \bibnamefont
  {Lubensky}},\ }\href {https://books.google.co.in/books?id=lAohAwAAQBAJ}
  {\emph {\bibinfo {title} {Principles of Condensed Matter Physics}}}\
  (\bibinfo  {publisher} {Cambridge University Press},\ \bibinfo {year}
  {2000})\BibitemShut {NoStop}%
\bibitem [{\citenamefont {Corral}\ \emph {et~al.}(2018)\citenamefont {Corral},
  \citenamefont {Sardany\'{e}s},\ and\ \citenamefont
  {Alsed\'a}}]{Alvaro_SciRep18}%
  \BibitemOpen
  \bibfield  {author} {\bibinfo {author} {\bibfnamefont {A.}~\bibnamefont
  {Corral}}, \bibinfo {author} {\bibfnamefont {J.}~\bibnamefont
  {Sardany\'{e}s}}, \ and\ \bibinfo {author} {\bibfnamefont {L.}~\bibnamefont
  {Alsed\'a}},\ }\href {\doibase 10.1038/s41598-018-30136-y} {\bibfield
  {journal} {\bibinfo  {journal} {Scientific Reports}\ }\textbf {\bibinfo
  {volume} {8}},\ \bibinfo {pages} {11783} (\bibinfo {year}
  {2018})}\BibitemShut {NoStop}%
\bibitem [{\citenamefont {Mijatović}\ \emph {et~al.}(2021)\citenamefont
  {Mijatović}, \citenamefont {Jovković}, \citenamefont {Janićević},
  \citenamefont {Graovac},\ and\ \citenamefont
  {Spasojević}}]{Mijatovic_physA21}%
  \BibitemOpen
  \bibfield  {author} {\bibinfo {author} {\bibfnamefont {S.}~\bibnamefont
  {Mijatović}}, \bibinfo {author} {\bibfnamefont {D.}~\bibnamefont
  {Jovković}}, \bibinfo {author} {\bibfnamefont {S.}~\bibnamefont
  {Janićević}}, \bibinfo {author} {\bibfnamefont {S.}~\bibnamefont
  {Graovac}}, \ and\ \bibinfo {author} {\bibfnamefont {D.}~\bibnamefont
  {Spasojević}},\ }\href {\doibase
  https://doi.org/10.1016/j.physa.2021.125883} {\bibfield  {journal} {\bibinfo
  {journal} {Physica A: Statistical Mechanics and its Applications}\ }\textbf
  {\bibinfo {volume} {572}},\ \bibinfo {pages} {125883} (\bibinfo {year}
  {2021})}\BibitemShut {NoStop}%
\bibitem [{\citenamefont {Mijatovi\ifmmode~\acute{c}\else \'{c}\fi{}}\ \emph
  {et~al.}(2019)\citenamefont {Mijatovi\ifmmode~\acute{c}\else \'{c}\fi{}},
  \citenamefont {Jovkovi\ifmmode~\acute{c}\else \'{c}\fi{}}, \citenamefont
  {Jani\ifmmode \acute{c}\else \'{c}\fi{}evi\ifmmode~\acute{c}\else
  \'{c}\fi{}},\ and\ \citenamefont {Spasojevi\ifmmode~\acute{c}\else
  \'{c}\fi{}}}]{Mijatovic_pre19}%
  \BibitemOpen
  \bibfield  {author} {\bibinfo {author} {\bibfnamefont {S.}~\bibnamefont
  {Mijatovi\ifmmode~\acute{c}\else \'{c}\fi{}}}, \bibinfo {author}
  {\bibfnamefont {D.}~\bibnamefont {Jovkovi\ifmmode~\acute{c}\else
  \'{c}\fi{}}}, \bibinfo {author} {\bibfnamefont {S.}~\bibnamefont
  {Jani\ifmmode \acute{c}\else \'{c}\fi{}evi\ifmmode~\acute{c}\else
  \'{c}\fi{}}}, \ and\ \bibinfo {author} {\bibfnamefont {D.}~\bibnamefont
  {Spasojevi\ifmmode~\acute{c}\else \'{c}\fi{}}},\ }\href {\doibase
  10.1103/PhysRevE.100.032113} {\bibfield  {journal} {\bibinfo  {journal}
  {Phys. Rev. E}\ }\textbf {\bibinfo {volume} {100}},\ \bibinfo {pages}
  {032113} (\bibinfo {year} {2019})}\BibitemShut {NoStop}%
\bibitem [{\citenamefont {Ahrens}\ and\ \citenamefont
  {Hartmann}(2011)}]{Ahrens_prb11}%
  \BibitemOpen
  \bibfield  {author} {\bibinfo {author} {\bibfnamefont {B.}~\bibnamefont
  {Ahrens}}\ and\ \bibinfo {author} {\bibfnamefont {A.~K.}\ \bibnamefont
  {Hartmann}},\ }\href {\doibase 10.1103/PhysRevB.83.014205} {\bibfield
  {journal} {\bibinfo  {journal} {Phys. Rev. B}\ }\textbf {\bibinfo {volume}
  {83}},\ \bibinfo {pages} {014205} (\bibinfo {year} {2011})}\BibitemShut
  {NoStop}%
\bibitem [{\citenamefont {Frontera}\ and\ \citenamefont
  {Vives}(1999)}]{Vives_pre99}%
  \BibitemOpen
  \bibfield  {author} {\bibinfo {author} {\bibfnamefont {C.}~\bibnamefont
  {Frontera}}\ and\ \bibinfo {author} {\bibfnamefont {E.}~\bibnamefont
  {Vives}},\ }\href {\doibase 10.1103/PhysRevE.59.R1295} {\bibfield  {journal}
  {\bibinfo  {journal} {Phys. Rev. E}\ }\textbf {\bibinfo {volume} {59}},\
  \bibinfo {pages} {R1295} (\bibinfo {year} {1999})}\BibitemShut {NoStop}%
\bibitem [{\citenamefont {Fytas}\ and\ \citenamefont
  {Mart\'{\i}n-Mayor}(2013)}]{Fytas_prl13}%
  \BibitemOpen
  \bibfield  {author} {\bibinfo {author} {\bibfnamefont {N.~G.}\ \bibnamefont
  {Fytas}}\ and\ \bibinfo {author} {\bibfnamefont {V.}~\bibnamefont
  {Mart\'{\i}n-Mayor}},\ }\href {\doibase 10.1103/PhysRevLett.110.227201}
  {\bibfield  {journal} {\bibinfo  {journal} {Phys. Rev. Lett.}\ }\textbf
  {\bibinfo {volume} {110}},\ \bibinfo {pages} {227201} (\bibinfo {year}
  {2013})}\BibitemShut {NoStop}%
\bibitem [{\citenamefont {Fytas}\ and\ \citenamefont
  {Mart\'{\i}n-Mayor}(2016)}]{Fytas_pre16}%
  \BibitemOpen
  \bibfield  {author} {\bibinfo {author} {\bibfnamefont {N.~G.}\ \bibnamefont
  {Fytas}}\ and\ \bibinfo {author} {\bibfnamefont {V.}~\bibnamefont
  {Mart\'{\i}n-Mayor}},\ }\href {\doibase 10.1103/PhysRevE.93.063308}
  {\bibfield  {journal} {\bibinfo  {journal} {Phys. Rev. E}\ }\textbf {\bibinfo
  {volume} {93}},\ \bibinfo {pages} {063308} (\bibinfo {year}
  {2016})}\BibitemShut {NoStop}%
\bibitem [{\citenamefont {P\'erez-Reche}\ \emph
  {et~al.}(2004{\natexlab{b}})\citenamefont {P\'erez-Reche}, \citenamefont
  {Tadi\ifmmode~\acute{c}\else \'{c}\fi{}}, \citenamefont {Ma\~nosa},
  \citenamefont {Planes},\ and\ \citenamefont {Vives}}]{Francisco_prl04}%
  \BibitemOpen
  \bibfield  {author} {\bibinfo {author} {\bibfnamefont {F.-J.}\ \bibnamefont
  {P\'erez-Reche}}, \bibinfo {author} {\bibfnamefont {B.}~\bibnamefont
  {Tadi\ifmmode~\acute{c}\else \'{c}\fi{}}}, \bibinfo {author} {\bibfnamefont
  {L.}~\bibnamefont {Ma\~nosa}}, \bibinfo {author} {\bibfnamefont
  {A.}~\bibnamefont {Planes}}, \ and\ \bibinfo {author} {\bibfnamefont
  {E.}~\bibnamefont {Vives}},\ }\href {\doibase 10.1103/PhysRevLett.93.195701}
  {\bibfield  {journal} {\bibinfo  {journal} {Phys. Rev. Lett.}\ }\textbf
  {\bibinfo {volume} {93}},\ \bibinfo {pages} {195701} (\bibinfo {year}
  {2004}{\natexlab{b}})}\BibitemShut {NoStop}%
\bibitem [{\citenamefont {Spasojevi\ifmmode~\acute{c}\else \'{c}\fi{}}\ \emph
  {et~al.}(2011)\citenamefont {Spasojevi\ifmmode~\acute{c}\else \'{c}\fi{}},
  \citenamefont {Jani\ifmmode \acute{c}\else
  \'{c}\fi{}evi\ifmmode~\acute{c}\else \'{c}\fi{}},\ and\ \citenamefont
  {Kne\ifmmode \check{z}\else \v{z}\fi{}evi\ifmmode~\acute{c}\else
  \'{c}\fi{}}}]{Spasojevic_prl11}%
  \BibitemOpen
  \bibfield  {author} {\bibinfo {author} {\bibfnamefont {D.}~\bibnamefont
  {Spasojevi\ifmmode~\acute{c}\else \'{c}\fi{}}}, \bibinfo {author}
  {\bibfnamefont {S.}~\bibnamefont {Jani\ifmmode \acute{c}\else
  \'{c}\fi{}evi\ifmmode~\acute{c}\else \'{c}\fi{}}}, \ and\ \bibinfo {author}
  {\bibfnamefont {M.}~\bibnamefont {Kne\ifmmode \check{z}\else
  \v{z}\fi{}evi\ifmmode~\acute{c}\else \'{c}\fi{}}},\ }\href {\doibase
  10.1103/PhysRevLett.106.175701} {\bibfield  {journal} {\bibinfo  {journal}
  {Phys. Rev. Lett.}\ }\textbf {\bibinfo {volume} {106}},\ \bibinfo {pages}
  {175701} (\bibinfo {year} {2011})}\BibitemShut {NoStop}%
\bibitem [{\citenamefont {Goldenfeld}(2018)}]{PTBook_Goldenfeld}%
  \BibitemOpen
  \bibfield  {author} {\bibinfo {author} {\bibfnamefont {N.}~\bibnamefont
  {Goldenfeld}},\ }\href {https://doi.org/10.1201/9780429493492} {\emph
  {\bibinfo {title} {Lectures on phase transitions and the renormalization
  group}}}\ (\bibinfo  {publisher} {CRC Press},\ \bibinfo {year}
  {2018})\BibitemShut {NoStop}%
\bibitem [{\citenamefont {Henkel}\ \emph {et~al.}(1998)\citenamefont {Henkel},
  \citenamefont {Andrieu}, \citenamefont {Bauer},\ and\ \citenamefont
  {Piecuch}}]{Henkel_prl98}%
  \BibitemOpen
  \bibfield  {author} {\bibinfo {author} {\bibfnamefont {M.}~\bibnamefont
  {Henkel}}, \bibinfo {author} {\bibfnamefont {S.}~\bibnamefont {Andrieu}},
  \bibinfo {author} {\bibfnamefont {P.}~\bibnamefont {Bauer}}, \ and\ \bibinfo
  {author} {\bibfnamefont {M.}~\bibnamefont {Piecuch}},\ }\href {\doibase
  10.1103/PhysRevLett.80.4783} {\bibfield  {journal} {\bibinfo  {journal}
  {Phys. Rev. Lett.}\ }\textbf {\bibinfo {volume} {80}},\ \bibinfo {pages}
  {4783} (\bibinfo {year} {1998})}\BibitemShut {NoStop}%
\bibitem [{\citenamefont {Ferdinand}\ and\ \citenamefont
  {Fisher}(1969)}]{Ferdinand_pr69}%
  \BibitemOpen
  \bibfield  {author} {\bibinfo {author} {\bibfnamefont {A.~E.}\ \bibnamefont
  {Ferdinand}}\ and\ \bibinfo {author} {\bibfnamefont {M.~E.}\ \bibnamefont
  {Fisher}},\ }\href {\doibase 10.1103/PhysRev.185.832} {\bibfield  {journal}
  {\bibinfo  {journal} {Phys. Rev.}\ }\textbf {\bibinfo {volume} {185}},\
  \bibinfo {pages} {832} (\bibinfo {year} {1969})}\BibitemShut {NoStop}%
\bibitem [{\citenamefont {Imry}\ and\ \citenamefont {Ma}(1975)}]{Imry_prl75}%
  \BibitemOpen
  \bibfield  {author} {\bibinfo {author} {\bibfnamefont {Y.}~\bibnamefont
  {Imry}}\ and\ \bibinfo {author} {\bibfnamefont {S.-k.}\ \bibnamefont {Ma}},\
  }\href {\doibase 10.1103/PhysRevLett.35.1399} {\bibfield  {journal} {\bibinfo
   {journal} {Phys. Rev. Lett.}\ }\textbf {\bibinfo {volume} {35}},\ \bibinfo
  {pages} {1399} (\bibinfo {year} {1975})}\BibitemShut {NoStop}%
\bibitem [{\citenamefont {Berthier}\ \emph {et~al.}(2020)\citenamefont
  {Berthier}, \citenamefont {Charbonneau},\ and\ \citenamefont
  {Kundu}}]{Berthier_prl20}%
  \BibitemOpen
  \bibfield  {author} {\bibinfo {author} {\bibfnamefont {L.}~\bibnamefont
  {Berthier}}, \bibinfo {author} {\bibfnamefont {P.}~\bibnamefont
  {Charbonneau}}, \ and\ \bibinfo {author} {\bibfnamefont {J.}~\bibnamefont
  {Kundu}},\ }\href {\doibase 10.1103/PhysRevLett.125.108001} {\bibfield
  {journal} {\bibinfo  {journal} {Phys. Rev. Lett.}\ }\textbf {\bibinfo
  {volume} {125}},\ \bibinfo {pages} {108001} (\bibinfo {year}
  {2020})}\BibitemShut {NoStop}%
\bibitem [{\citenamefont {Ray}\ and\ \citenamefont {Klein}(1990)}]{Ray_jsp90}%
  \BibitemOpen
  \bibfield  {author} {\bibinfo {author} {\bibfnamefont {T.}~\bibnamefont
  {Ray}}\ and\ \bibinfo {author} {\bibfnamefont {W.}~\bibnamefont {Klein}},\
  }\href {\doibase 10.1007/BF01027308} {\bibfield  {journal} {\bibinfo
  {journal} {J. Stat. Phys.}\ }\textbf {\bibinfo {volume} {61}},\ \bibinfo
  {pages} {891} (\bibinfo {year} {1990})}\BibitemShut {NoStop}%
\bibitem [{\citenamefont {Ray}(1991)}]{Ray_jsp91}%
  \BibitemOpen
  \bibfield  {author} {\bibinfo {author} {\bibfnamefont {T.}~\bibnamefont
  {Ray}},\ }\href {\doibase 10.1007/BF01020882} {\bibfield  {journal} {\bibinfo
   {journal} {J. Stat. Phys.}\ }\textbf {\bibinfo {volume} {62}},\ \bibinfo
  {pages} {463} (\bibinfo {year} {1991})}\BibitemShut {NoStop}%
\bibitem [{\citenamefont {Gagliardi}\ and\ \citenamefont
  {Macheda}(2021)}]{Gagliardi_pre21}%
  \BibitemOpen
  \bibfield  {author} {\bibinfo {author} {\bibfnamefont {G.}~\bibnamefont
  {Gagliardi}}\ and\ \bibinfo {author} {\bibfnamefont {F.}~\bibnamefont
  {Macheda}},\ }\href {\doibase 10.1103/PhysRevE.104.014115} {\bibfield
  {journal} {\bibinfo  {journal} {Phys. Rev. E}\ }\textbf {\bibinfo {volume}
  {104}},\ \bibinfo {pages} {014115} (\bibinfo {year} {2021})}\BibitemShut
  {NoStop}%
\bibitem [{\citenamefont {Liu}\ \emph {et~al.}(2016{\natexlab{b}})\citenamefont
  {Liu}, \citenamefont {Grinberg},\ and\ \citenamefont {Rappe}}]{Liu_nat16}%
  \BibitemOpen
  \bibfield  {author} {\bibinfo {author} {\bibfnamefont {S.}~\bibnamefont
  {Liu}}, \bibinfo {author} {\bibfnamefont {I.}~\bibnamefont {Grinberg}}, \
  and\ \bibinfo {author} {\bibfnamefont {A.~M.}\ \bibnamefont {Rappe}},\ }\href
  {\doibase 10.1038/nature18286} {\bibfield  {journal} {\bibinfo  {journal}
  {Nature}\ }\textbf {\bibinfo {volume} {534}},\ \bibinfo {pages} {360}
  (\bibinfo {year} {2016}{\natexlab{b}})}\BibitemShut {NoStop}%
\bibitem [{\citenamefont {Cohen}(1992)}]{Cohen_nat92}%
  \BibitemOpen
  \bibfield  {author} {\bibinfo {author} {\bibfnamefont {R.~E.}\ \bibnamefont
  {Cohen}},\ }\href {\doibase 10.1038/358136a0} {\bibfield  {journal} {\bibinfo
   {journal} {Nature}\ }\textbf {\bibinfo {volume} {358}},\ \bibinfo {pages}
  {136} (\bibinfo {year} {1992})}\BibitemShut {NoStop}%
\bibitem [{\citenamefont {Shekhawat}\ \emph {et~al.}(2013)\citenamefont
  {Shekhawat}, \citenamefont {Zapperi},\ and\ \citenamefont
  {Sethna}}]{Shekhawat_prl13}%
  \BibitemOpen
  \bibfield  {author} {\bibinfo {author} {\bibfnamefont {A.}~\bibnamefont
  {Shekhawat}}, \bibinfo {author} {\bibfnamefont {S.}~\bibnamefont {Zapperi}},
  \ and\ \bibinfo {author} {\bibfnamefont {J.~P.}\ \bibnamefont {Sethna}},\
  }\href {\doibase 10.1103/PhysRevLett.110.185505} {\bibfield  {journal}
  {\bibinfo  {journal} {Phys. Rev. Lett.}\ }\textbf {\bibinfo {volume} {110}},\
  \bibinfo {pages} {185505} (\bibinfo {year} {2013})}\BibitemShut {NoStop}%
\end{thebibliography}%
\end{document}